\pdfoutput=1

\documentclass[a4paper,11pt]{article}
\usepackage{jcappub}

\usepackage{graphicx}
\usepackage{placeins}

\usepackage{xcolor}
\usepackage{amsmath}
\usepackage{amssymb}
\usepackage{bm}
\usepackage{slashed}
\usepackage{xspace}
\usepackage{makecell}  
\usepackage{adjustbox} 
\usepackage{subcaption}
\usepackage{caption}
\usepackage{enumitem} 
\usepackage{mathtools}
\usepackage{chngcntr}
\usepackage{aas_macros}
\usepackage[normalem]{ulem}
\usepackage{fontawesome5}
\usepackage{hyperref}
\usepackage{orcidlink}

\usepackage{multirow}
\usepackage{array}
\newcolumntype{L}[1]{>{\raggedright\let\newline\\\arraybackslash\hspace{0pt}}m{#1}}
\newcolumntype{C}[1]{>{\centering\let\newline\\\arraybackslash\hspace{0pt}}m{#1}}
\newcolumntype{R}[1]{>{\raggedleft\let\newline\\\arraybackslash\hspace{0pt}}m{#1}}
\title{Inferring cosmological parameters from galaxy and dark sirens cross-correlation}

\author[a]{Giona~Sala\orcidlink{0009-0001-3716-862X},}
\affiliation[a]{Institute for Theoretical Particle Physics and Cosmology, RWTH Aachen University, Sommerfeldstr. 16, D-52056 Aachen, Germany}
\emailAdd{gsala@physik.rwth-aachen.de}

\author[b,c]{Alessandro~Cuoco\orcidlink{0000-0003-1504-894X},}
\affiliation[b]{Department of Physics, University of Torino, via P. Giuria 1, 10125 Torino, Italy}
\affiliation[c]{
Istituto Nazionale di Fisica Nucleare, Sezione di Torino, via P. Giuria 1, 10125 Torino, Italy}

\author[a]{Julien~Lesgourgues\orcidlink{0000-0001-7627-353X},}

\author[d]{Konstantinos-Rafail~Revis\orcidlink{0000-0002-5645-5448},}
\affiliation[d]{Department of Computer Science, Paderborn University, Warburger Str. 100, 33098, Paderborn, Germany}

\author[e,f]{Lorenzo~Valbusa~Dall'Armi\orcidlink{0000-0002-0412-8058}}
\affiliation[e]{Dipartimento di Fisica ``Enrico Fermi'', Università di Pisa, Largo Bruno Pontecorvo 3, Pisa I-56127, Italy}
\affiliation[f]{INFN, Sezione di Pisa, Largo Bruno Pontecorvo 3, Pisa I-56127, Italy}

\author[a,g]{and~Santiago~Casas\orcidlink{0000-0002-4751-5138}}
\affiliation[g]{Institute of Scientific Information, German Aerospace Center (DLR), Linder Höhe, D-51147 Köln, Germany}

\abstract{The number of observed gravitational wave (GW) events is growing fast thanks to rapidly improving detector sensitivities. GWs from compact binary coalescences like Black Holes or Neutron Stars behave like standard sirens and can be used as cosmological probes. To this aim, generally, the observation of an electromagnetic counterpart and the measurement of the redshift are needed. However, even when those are not available, it is still possible to exploit these ``dark sirens'' via statistical methods. In this work, we explore a method that exploits the information contained in the cross-correlation of samples of GW events with matter over-density tracers like galaxy catalogues. Contrary to other currently employed dark-sirens methods, this approach does not suffer from systematic errors related to the incompleteness of the galaxy catalogue. To further enhance the technique, we implement tomography in redshift space for the galaxy catalogue and luminosity distance space for the GWs. We simulate future data collected by the array of currently existing 
detectors, namely LIGO, Virgo, and KAGRA, as well as planned third-generation ones such as the Einstein Telescope and Cosmic Explorers. We cross-correlate these data with those from upcoming photometric galaxy surveys such as Euclid. We perform a sensitivity forecast employing a full-likelihood approach and explore the parameter space with Monte Carlo Markov Chains. We find that with this method, third-generation detectors will be able to determine the Hubble constant $H_0$ with an error of only $0.7\%$, 
which is enough to provide decisive information to shed light on the Hubble tension.
Furthermore, for the other cosmological parameters, we find that the GWs and galaxy surveys information are highly complementary, and the use of both significantly improves the ability to constrain the underlying cosmology.
}

\begin{document} 
\maketitle
\flushbottom


\section{Introduction}

The detection of Gravitational Waves (GW) \cite{LIGOScientific:2016aoc} has opened a new era in cosmology and astrophysics. The events observed up to now result from a merging of two massive compact objects in a close orbit, i.e., Compact Binary Coalescences (CBC),
and the number of detections is constantly increasing \cite{GWTC4_1,GWTC4_2}.
So far,
only one merger of two binary neutron stars (BNSs) has been detected.
\cite{LIGOScientific:2017vwq}, while the majority is the result of binary black hole (BBH) mergers \cite{LIGOScientific:2020kqk,KAGRA:2021vkt,KAGRA:2021duu}. 
The 
third type of event
consists of a binary black hole neutron star merger (BHNS)  \cite{BHNSoverview,LIGOScientific:2021qlt}. The observed BHs are believed to be of stellar origin, although it is not excluded that a small fraction could be primordial \cite{raidal2017,PBH2023,bagui2023,raidal2024}.

CBCs behave as standard sirens, and thus they provide a measurement of the luminosity distance $D_L$ of the event. If the redshift of the event can be independently measured, the Hubble diagram can be built, from which cosmological inference can be performed \cite{Jin:2025dvf,Pierra:2025fgr}. This method, dubbed the bright standard sirens method, has been possible so far only for the single case of the BNS merger GW170817 \cite{LIGOScientific:2017vwq,LIGOScientific:2017adf}, providing a $\sim \! 10\%$ determination on the Hubble constant $H_0$.  
BBHs, on the other hand, are unlikely to produce an electromagnetic counterpart from which to measure the redshift 
(see \cite{EM_BBH_2023} for a search of an electromagnetic signal from BBH events).  
And even for BNSs,
the expected electromagnetic counterparts will typically be very faint and hard to detect for most of the events. As a result, most of the detected GWs will thus remain dark and without a measured redshift. 
Fortunately, however, it is still possible 
to exploit these dark sirens for cosmology using statistical methods, even if the redshift of a single event is not available. These methods are based on the fact that matter in the universe is not homogeneous and isotropic but clusters along the so-called Large Scale Structures (LSS) of the Universe, with CBCs also following the LSS.
So far, this method has involved the use of an auxiliary galaxy catalogue (typically the Glade+ catalogue \cite{Dalya:2021ewn}) which is built in such a way as to contain the galaxies likely to host the GW events. Thus, for a specific GW event with a direction and luminosity distance known up to some reconstruction errors, one can look for the possible host galaxies within the catalogue and assign a statistical redshift to the event. Applying this method to the currently available dataset of about few dozen GW events does not provide significant constraints on $H_0$, while, in conjunction with GW170817, it provides a minor improvement of the constraint \cite{Gray:2023wgj,LIGOScientific:2019zcs,LIGOScientific:2021aug, Beirnaert2025,Gair:2022zsa, Borghi_2024}. Forecasts with future datasets are, however, promising 
\cite{Zhu:2021aat,Ghosh2024,ghosh2025}.

A problem of the above method is that it is prone to the choice of the specific catalogue used, and in particular, to its completeness in terms of the galaxies that indeed host the GW events, a systematic that is very difficult to assess and quantify.
Here, we propose an alternative novel method, which is very robust and does not suffer from the above systematic, and which is based on the cross-correlation of catalogues of GW events and tracers of the LSS, like, e.g., galaxy catalogues as the Dark Energy Spectroscopic Instrument (DESI) \cite{DESI:2016fyo}, the Spectro-Photometer for the History of the Universe, Epoch of Reionization, and Ices Explorer (SPHEREx) \cite{Dore:2018kgp},  Euclid \cite{Euclid:2019clj,Euclid:Mellier2024} and the Large Synoptic Survey Telescope (LSST) \cite{LSSTScience:2009jmu,LSST:2008ijt}.
We enhance this technique with the use of tomography in redshift space for the galaxy catalogues and luminosity distance space for the GW catalogues. The use of tomography is crucial to optimally exploit the information and to increase the sensitivity of the analysis.
To implement this method, we partly employ the formalism developed in previous works in the literature  \cite{Namikawa:2016edr,Oguri:2016dgk,Camera:2013xfa, Namikawa:2015prh,Scelfo:2018sny, Raccanelli:2016cud, Mukherjee:2018ebj, Scelfo:2021fqe,Mukherjee:2020hyn, Diaz:2021pem,Libanore:2021jqv,Libanore:2020fim,Calore:2020bpd,Semenzato_2024}, and expand it for the present purpose. We then use the formalism to perform sensitivity forecasts on cosmological parameters with simulations of future GW data from the current, second generation (2G), of GW detectors, namely the Laser Interferometer Gravitational-Wave Observatory (LIGO) \cite{LIGOScientific:2014pky}, VIRGO \cite{VIRGO:2014yos} and the Kamioka Gravitational Wave Detector (KAGRA) \cite{KAGRA:2020tym}, as well as from the planned third generation (3G) detectors Einstein Telescope (ET)\cite{Hild:2010id,ET,Punturo:2010zz} and Cosmic Explorers (CE) \cite{CE}.

An application of this formalism for cosmology inference has also recently been discussed in \cite{Ferri_2025, pedrotti2025}. 
The main difference in our approach compared to \cite{pedrotti2025} is the use of a full likelihood formalism instead of a Fisher matrix to perform the forecast. Furthermore, we also forecast  the sensitivity achievable with the current generation of GW detectors, while \cite{pedrotti2025} only focuses on forecasts for the future 3G detectors. With respect to \cite{Ferri_2025}, we instead improve the analysis by enlarging the cosmological parameters space explored and studying the interplay and complementarity between GWs and galaxy surveys observables. Finally, we perform a forecast using both BBHs and BNSs, while \cite{Ferri_2025, pedrotti2025} only include BBHs.
To implement this full-likelihood analysis, we run Monte Carlo Markov Chains (MCMC) scans of the parameter space with the code \texttt{MontePython} \href{https://github.com/brinckmann/montepython_public}{\faGithub} \cite{Brinckmann:2018cvx, Audren:2012wb}, which in turn uses \texttt{CLASS} \href{https://github.com/lesgourg/class_public}{\faGithub} \cite{CLASS1,CLASS2,CLASS3,CLASS4} for theoretical predictions of distances and power spectra.

The article is structured as follows.
In \autoref{sec:power} we introduce and detail the formalism used to predict the angular power spectra of cross-correlation between matter overdensity tracers and GWs. In \autoref{sec:GW} we describe the analysis employed to simulate future realistic catalogues of GWs from various GW detector configurations. In 
\autoref{sec:likelihood} we discuss the likelihood formalism, while in \autoref{sec:results} we show the results of our sensitivity forecasts on cosmological parameters, in particular $H_0$, from the cross-correlation method. Finally, we present our conclusions in \autoref{sec:conclusion}.

In our analysis, we use natural units with $c=\hbar=1$.

\vspace{2cm}

\section{Power spectra $C_\ell$'s}\label{sec:power}

We treat galaxies and dark sirens as two linear biased tracers of the underlying matter density field, such that in Fourier space and on the large cosmological scales of interest for this work,
\begin{equation}
    \centering\label{eq:bias_concept}
    \delta^{A}(\vec{k},z) = b^{A}(z) \, \delta_m(\vec{k},z)~,
\end{equation}
where $\delta^{A}$ stands for the relative overdensity of galaxies for $A=G$ and of gravitational wave sources for $A=W$, 
$\delta_m$ is the total matter overdensity, and $b^{A}(z)$ is the $z$-dependent bias of probe $A$. Using the Limber approximation, the auto-correlation and cross-correlation power spectra of two linear tracers $A$ and $B$ can be expressed in terms of multipoles $C_\ell$ as \cite{Calore:2020bpd,EUCLID:2023uep}:
\begin{equation}
    \centering\label{eq:C_general}
     C^{A_iB_j}(\ell) = \int_{z_{\rm min}}^{z_{\rm max}}dz\frac{{W^{A_i}(z)}W^{B_j}(z)}{H(z)r^2(z)}\times P_{\delta\delta}\left[k=\frac{\ell+1/2}{r(z)},z\right] ,
\end{equation} 
where we have introduced a tomographic approach by breaking down the distribution of each source into redshift bins. 
Large catalogues, combined with precise distance measurements, allow observables to be partitioned into distance bins, effectively serving as independent probes and enhancing the inference, as shown in \cite{Calore:2020bpd}. This tomographic approach is crucial for extracting maximum information and deriving optimal constraints from a likelihood analysis\footnote{The exact number of bins used in the computations is discussed in \autoref{sec:likelihood}.}.
In what follows, indices $i$, $j$, $k$, $n$ label the specific bins, while $A$, $B$, $C$, $D$ are used for the source population, either $G$ for galaxies or $W$ for GW sources. 
Furthermore, in \eqref{eq:C_general},
$\ell$ is the multipole, $k$ the comoving wavenumber, $r(z)$ the comoving distance at redshift $z$, $H(z)$ the expansion rate at $z$, $W^{A_i}(z)$ the window function of the $A$ observable in the $i$-th redshift bin, and $P_{\delta\delta}(k, z)$ the non-linear matter power spectrum.  
The liner power spectrum is calculated with \texttt{CLASS} and the non-linear correction using the \texttt{halofit} \cite{Takahashi2012} option.
These functions depend on the underlying cosmology and its defining parameters.
We will consider the following parameters:
 baryon density $\omega_{\rm b}$,
 cold dark matter density $\omega_{\rm cdm}$,
scalar spectral index $n_{\rm s}$,
scalar amplitude $A_{\rm s}$ and
 Hubble parameter $h$, as listed later in \autoref{tab:cosmo_nuisance}.
We adopt units such that $C^{A_iB_j}(\ell)$ is dimensionless. The integral boundaries $z_{\rm min}$ and $z_{\rm max}$ account for the redshift range covered by the surveys.
The window functions can be explicitly written as: 
\begin{equation}
    \centering
    \label{eq:weight_func_def}
    W^{A_i}(z) = \frac{dn^{A_i}}{dz}\,\,\frac{dz}{dr}\,\,b^A(z) = \frac{dn^{A_i}}{dz}\,\,H(z)\,\,b^A(z)~,
\end{equation}
where $\frac{dn^{A_i}}{dz}$ represents the unit-normalized redshift distribution of population $A$ in the specific redshift bin $i$. Furthermore, $\frac{dn^{A_i}}{dz}$ takes into account the effect of uncertainties in the measured distance or redshift to the source, as discussed below.
There is a crucial difference regarding the definition of this quantity for the two tracers.
In galaxy surveys, one directly measures galaxy redshifts, and the window function follows directly from \eqref{eq:weight_func_def},
\begin{equation}
    \centering
    \label{eq:window_func_G}
    W^{G_i}(z) =\frac{dn^{G_i}}{dz}(z)\,\, H(z)\,\,b^{G}(z)~.
\end{equation}
On the other hand, in GW surveys, one only gets an estimate of the luminosity distance $D_L$ to the source.
Thus, the window function can be expressed as
\begin{equation}
    \centering
    \label{eq:weight_func_W}
    W^{W_i}(z) 
    = \frac{dn^{W_i}}{dD_L}(D_L(z))\,\, \frac{dD_L}{dz}(z)\,\, 
  H(z)\,\, b^W(z)~.
\end{equation}
Using the definition of $D_L$ in a flat universe, one gets
\begin{equation}\centering\label{eq:dDLdz_flat}
    \frac{dD_L}{dz}(z)=\frac{D_L(z)}{1+z}+\frac{1+z}{H(z)}~.
\end{equation}
This is straightforward to generalize to the case of a spatially curved universe (see \autoref{app:dl_z}).

Regarding galaxies, we will use the specifics of upcoming photometric galaxy catalogues similar to Euclid \cite{2011arXiv1110.3193L,refId0} or LSST \cite{LSST:2008ijt,LSSTScience:2009jmu}. Photometric catalogues are expected to provide roughly 5\% accuracy in redshift determination and a sample of about 2 billion galaxies, and are ideal for cross-correlation with GWs because of the very large number of galaxies and resulting low shot-noise. 
Spectroscopic samples have a much better redshift determination, but a lower number of galaxies and large shot-noise. The performance of this analysis with a spectroscopic catalogue, in comparison to photometric surveys, is non-trivial and would require a dedicate study to be assessed which we leave for future works.

\begin{figure}[t]
    \centering
    \includegraphics[width=0.7\textwidth]{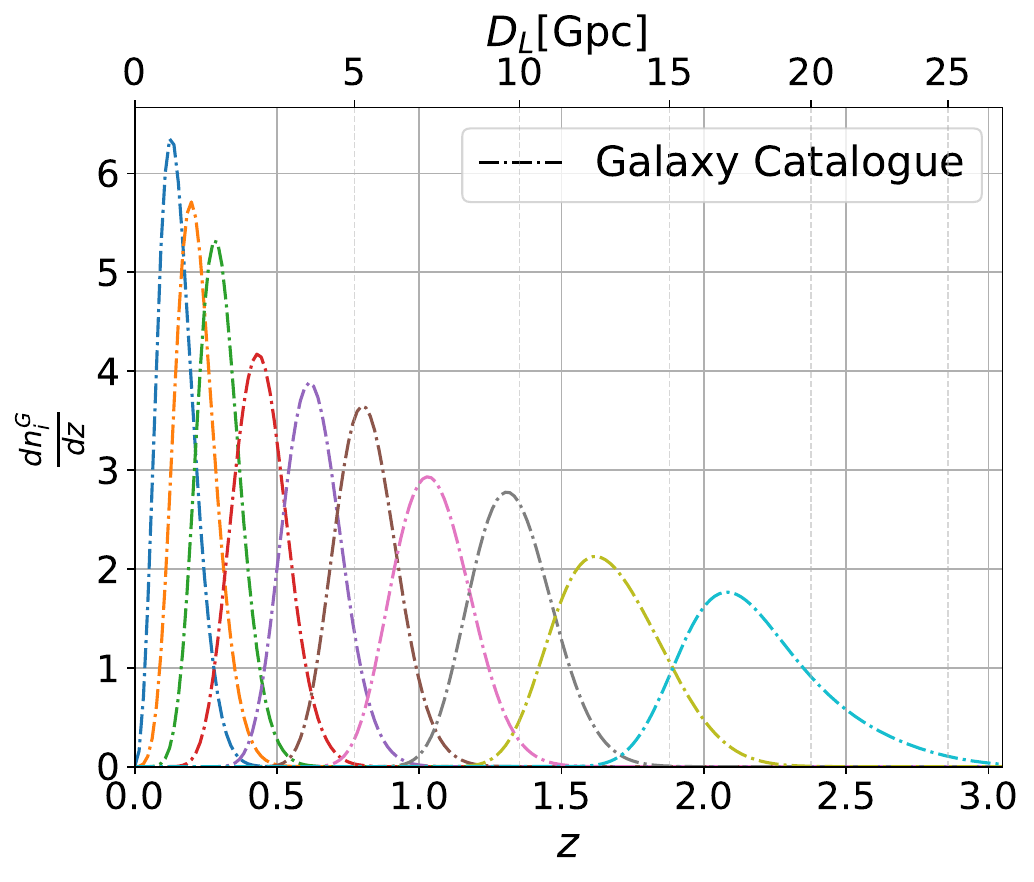}
    \caption{For the galaxy redshift survey assumed in our forecast, normalised density distribution $dn^{G_i}/dz$ as a function of $z$ for each of our 10 $z$-bins.
    The upper $x$-axis shows the corresponding luminosity distance $D_L$ according to the fiducial cosmology.}
    \label{fig:bin_dist_G}
\end{figure}

As mentioned above, the quantities $\frac{dn^{G_i}}{dz}$ and $\frac{dn^{W_i}}{dD_L}$ include the effect of redshift or distance measurement errors. 
Regarding $\frac{dn^{G_i}}{dz}$, in the case of a Euclid-like survey, this amounts in multiplying\footnote{Note that $\frac{dN^G}{dz}(z)$ involves $z$ and not $z'$ in \eqref{eq:conv_n_i_G}, i.e., this function can be, in principle, moved outside of the integral.
In principle, the difference with respect to the standard definition where the prime is present is small if the bin is small enough, since $\frac{dN^G}{dz}(z)$ would be roughly constant within this bin. We have verified that this is indeed the case except for bins at low redshift below 0.2, where some sizeable difference appears. Nonetheless, this definition has become customary in recent papers (see e.g. \cite{Euclid:2019clj,EUCLID:2023uep}), so we stick to the Euclid convention.}
the true underlying source distribution $\frac{dN^G}{dz}(z)$ with the photometric redshift error function $p_{\rm ph}^{G}(z',z)$ of the galaxy survey \cite{Euclid:2022oea, EUCLID:2023uep, Euclid:2019clj, Chevallier:2000qy, Busch2020},
\begin{equation}
\centering\label{eq:conv_n_i_G}
\frac{dn^{G_i}}{dz}(z)=\frac{\int_{z_i^{-}}^{z_i^{+}}dz' \,\, \frac{dN^G}{dz}(z) \,\, p_{\rm ph}^{G}(z',z)}{\int_{z_{\rm min}}^{z_{\rm max}}dz\int_{z_i^{-}}^{z_i^{+}}dz' \,\, \frac{dN^G}{dz}(z) \,\, p_{\rm ph}^{G}(z',z)},
\end{equation}
where $z_i^{-}$ and $z_i^{+}$ are the edges of the $i$-th redshift bin, and $z_{\rm min}$ and $z_{\rm max}$ are the edges of the survey. The denominator ensures that the distribution is unit-normalised in each bin. For the true underlying galaxy distribution, we assume the form commonly employed for a 
Euclid-like survey\footnote{Whilst more realistic galaxy distributions for photometric surveys exist \cite{2019Kitching,2021EuclidXII}, in this work, we make use of the phenomenological approximation from the Euclid Red Book \cite{2011arXiv1110.3193L}. This approach is widely used in Euclid-like analyses, and it has been verified that it does not significantly influence the results \cite{Euclid:2019clj,Euclid:2022oea}.}
\cite{2011arXiv1110.3193L,Euclid:2019clj,Euclid:2022oea,EUCLID:2023uep},
\begin{equation}
    \centering\label{eq:n_g}
    \frac{dN^G}{dz}(z) = \left(\frac{z}{z_0}\right)^2\exp\left[-\left(\frac{z}{z_0}\right)^{3/2}\right]~,
\end{equation}
with $z_0 = z_{\rm mean}/\sqrt{2}$ and $z_{\rm mean} = 0.9$ according to \cite{EUCLID:2023uep}. The distribution is normalised to give a total of $1.6 \times 10^9$ galaxies within the survey area, which covers $f_{\rm fov}^{G}\approx 0.3636$ of the full sky. This corresponds to a galaxy density of about 30 arcmin$^{-2}$ \cite{EUCLID:2023uep}.
The photometric redshift error function is parameterised as the sum of two Gaussians, the second one modelling a fraction $f_{\text{out}}$ of catastrophic outliers,
\begin{align}
    \centering\label{eq:G_error}
    p_{ph}^{G}(z', z) =& \frac{1-f_{\text{out}}}{\sqrt{2\pi}\sigma_{b}(1+z)} \exp\left(-\frac{(z-z')^2}{2\sigma_{b}^2(1+z)^2}\right) 
    \\\nonumber
    &+ \frac{f_{\text{out}}}{\sqrt{2\pi}\sigma_{\rm out}(1+z)} \exp\left(-\frac{(z-z'-z_{\rm out})^2}{2\sigma_{\rm out}^2(1+z)^2}\right) ~.
\end{align}
Following \cite{EUCLID:2023uep}, we take $\sigma_{\rm out} =\sigma_b= 0.05$, standing for a constant relative redshift error of 5\%, $z_{\rm out}=0.1$, and finally $f_{\text{out}} = 0.1$, that is, a 10\% fraction of outliers.
Considering the above 5\% redshift error, we divide the redshift range into 10 bins with the following edges: $[0.0, 0.1, 0.2, 0.3, 0.5, 0.7, 0.9, 1.2, 1.5, 2.0, 3.0]$. The resulting galaxy distribution $\frac{dn_i^G}{dz}(z)$ in each bin is shown in \autoref{fig:bin_dist_G}.

Regarding $\frac{dn^{W_i}}{dD_L}$, we convolve the true underling source distribution $\frac{dN^W}{dD_L}(D_L)$ with the luminosity distance error function $p_{\rm err}^{W_i}(D_L',D_L)$ of the GW catalogue,
\begin{equation}
\centering\label{eq:conv_n_i_W}
\frac{dn^{W_i}}{dD_L}(D_L)=\frac{\int_{z_i^{-}}^{z_i^{+}}dD_L' \,\, \frac{dN^W}{dD_L}(D_L') \,\, p_{\rm err}^{W_i}(D_L',D_L)}{\int_{D_{L{\rm min}}}^{D_{L{\rm max}}}dD_L\int_{D_{Li}^{-}}^{D_{Li}^{+}}dD_L' \,\, \frac{dN^G}{dz}(D_L') \,\, p_{\rm err}^{W_i}(D_L',D_L)},
\end{equation}
modelling the error function as a single Gaussian that accounts for a relative error $\delta\sigma_{D_L}=\frac{\sigma_{D_L}}{D_L}$,
\begin{equation}
    \centering\label{eq:W_error}
    p_{\rm err}^{W_i}\left(D_L', D_L\right) = \frac{1}{\delta\sigma_{D_L}^{W_i}\cdot D_L\sqrt{2\pi}} \exp\left(\frac{-\left(D_L-D_L'\right)^2}{2 \left(\delta\sigma_{D_L}\cdot D_L\right)^2}\right)~.
\end{equation}
The value of $\delta\sigma_{D_L}^{W_i}$ can depend on the luminosity distance bin and the type of GW catalogue used. Specific values appropriate for the various cases considered are discussed in \autoref{sec:GW}.
The true underlying distribution $\frac{dN^W(D_L)}{dD_L}$ will be also discussed in \autoref{sec:GW}. 

Note that for the $C_\ell$ calculation and in order to correlate the two tracers, both sources must be expressed as a function of the same variable $z$.
The relation between $D_L$ and $z$ is computed by \texttt{CLASS}\footnote{More details about this relation are given in \autoref{app:dl_z}.} for each assumed cosmology.

Regarding the bias, for galaxies, we assume the commonly employed form
\cite{Euclid:2022oea, Euclid:2019clj,  Busch2020}
\begin{equation}
    \centering\label{eq:bias_gc}
    b^{G}(z) = a_1^{G}\sqrt{1+z}~.
\end{equation}
The parameter $a_1^{G}$ will be considered a nuisance parameter in the likelihood analysis and left free to vary around the fiducial value $a_1^{G}=1$. Furthermore, for simplicity, we actually approximate $b^{G}(z)$ as constant within each $z$-bin and equal to the value at the center of the bin. For GWs, we adopt \cite{Diaz:2021pem}
\begin{equation}
        \centering\label{eq:bias_gw}
        b^{W}(z) = a_1^{W}(1+z)^{a_2^{W}}~,
\end{equation}
considering $a_1^{W}$ and $a_2^{W}$ as two nuisance parameters
with fiducial values $a_1^{W}=2$ and $a_2^{W}=0$. 
We also approximate $b^{W}(z)$ as constant within each $z$-bin and equal to the value at the center of the bin. 
Together with the cosmological parameters, these nuisance variables constitute the complete parameter set to be inferred, and are summarised in \autoref{tab:cosmo_nuisance}.

Finally, for the calculation of the error on the $C_\ell$s (see \autoref{sec:likelihood})  we will need  the noise spectrum 
\begin{equation}
    \centering\label{eq:noise_def}
    N^{A_iB_j}(\ell) = \frac{4\pi f_{\rm fov}^{A}}{N^{A_i}}
    \frac{\delta_{A_iB_j}}{(\mathcal{W}^{A_i}(\ell))^2}~,  
\end{equation}
which depends on $f_{\rm fov}^A$, the fraction of sky observed by tracer $A$, and $N^{A_i}$, the number of events in the $i$-th bin. We take $f_{\rm fov}^{W}=1$ for GWs and $f_{\rm fov}^{G}\approx 0.3636$ for galaxies \cite{EUCLID:2023uep}. The Kronecker symbol $\delta_{A_iB_j}$ accounts for the fact that, in general, there is no noise correlation between different bins or sources. The beam window function $\mathcal{W}_{i}^{A}(\ell)$ includes the angular localisation uncertainty of sources in the sky. Assuming a circular beam with a Gaussian profile, it is given by
\begin{equation}\centering\label{eq:beam_ang}
    \mathcal{W}^{A_i}(\ell)=\exp\left(-\frac{(\sigma^{A_i})^2 \ \ell^2}{2}\right)~,
\end{equation}
where $\sigma^{A_i}$ represents the beam width. 
For GW sources, we take $\sigma^{W_i}$ from \cite{Calore:2020bpd}, as discussed in more detail in \autoref{sec:GW}. GW localisation errors typically range from sub-degree for the best localised events to several degrees for the poorly localised ones. 
Instead, for galaxies, we take $\mathcal{W}^{G_i}(\ell)=1$ since the localisation error is of the order of a few arcseconds, and thus completely negligible for any multipole $\ell$ of interest.
Note that our definition of noise is unconventional, since the usual  definition typically does not include the beam window. Nonetheless, in our calculations, the noise and the beam always appear together, as in  \eqref{eq:noise_def}. So, for convenience, we adopt the above definition. The noise is non-null only for the galaxy auto-correlations and GW auto-correlations, where it reads:  
\begin{eqnarray}
    N^{G_iG_j} & = &  \frac{4\pi f_{\rm fov}^{G}}{N^{G_i}}\delta_{ij}~, \\
    N^{W_iW_j}(\ell) & = &  \frac{4\pi }{N^{W_i}}e^{(\sigma^{W_i})^2 \ell^2}\delta_{ij}~.
\end{eqnarray}

Combining all the elements discussed above and inserting them into \eqref{eq:C_general}, we obtain explicit expressions for our auto-correlation and cross-correlation $C_\ell$'s,
\begin{align}\centering\label{eq:Cl_GG}
C^{G_iG_j}(\ell) &= \int_{z_{\rm min}}^{z_{\rm max}} dz \frac{P_{\delta\delta} \left( \frac{\ell+1/2}{r(z)}, z \right)}{r(z)^2} H(z) \frac{dn^{G_i}(z)}{dz}  \frac{dn^{G_j}(z)}{dz} \left[ b^G(z) \right]^2,
\\
\label{eq:Cl_WG}
C^{W_iG_j}(\ell) &= C^{G_jW_i}(\ell) = \int_{z_{\rm min}}^{z_{\rm max}} dz \frac{P_{\delta\delta} \left( \frac{\ell+1/2}{r(z)}, z \right)}{r(z)^2} H(z)  \frac{dn^{W_i}(D_L(z))}{dD_L} 
\frac{dn^{G_j}(z)}{dz}
\nonumber \\
&~~~~~~~~~~~~~~~~~~~~~~~~~~~~~~~~\times \left( \frac{D_L(z)}{1+z} + \frac{1+z}{H(z)} \right) b^W(z) b^G(z)~,
\\
\label{eq:Cl_WW}
C^{W_iW_j}(\ell) &= \int_{z_{\rm min}}^{z_{\rm max}} dz\frac{P_{\delta\delta} \left( \frac{\ell+1/2}{r(z)}, z \right)}{r(z)^2} H(z) \frac{dn^{W_i}(D_L(z)) }{dD_L} \frac{dn^{W_j}(D_L(z)) }{dD_L}
\\\nonumber
&~~~~~~~~~~~~~~~~~~~~~~~~~~~~~~~~\times\left( \frac{D_L(z)}{1+z} + \frac{1+z}{H(z)} \right)^2 \left[ b^W(z) \right]^2
\end{align}
A selection of representative $C_\ell$s for our fiducial cosmology is shown in \autoref{app:cls}.

\section{Modelling of GW future data}\label{sec:GW}

We consider a forecast for 3 different GW detector configurations:
\begin{itemize}
    \item {\bf HLVK} corresponds to the four currently existing detectors: the Hanford and Livingstone LIGO detectors \cite{LIGOScientific:2014pky} in the USA, Virgo in Italy  \cite{VIRGO:2014yos} and KAGRA  \cite{KAGRA:2020tym} in Japan. For the forecast, we assume $10$~ years of data collection at full design sensitivity.

    \item {\bf HLVIK} includes the same detectors plus the planned LIGO India detector \cite{Unnikrishnan:2013qwa,Unnikrishnan:2023uou}. 
    
    \item {\bf ET2CE} stands for an array of three planned next-generation (3G) detectors, consisting of the Europe-based Einstein Telescope (ET) \cite{ET} with a triangular configuration and the two interferometers of the US-based Cosmic Explorers (CE) \cite{CE} collaboration.
\end{itemize}
As GW populations, we will consider only BBH and BNS. A population of mixed BHNS is also expected. Two events of this type have been detected so far \cite{LIGOScientific:2021qlt}. However, given the still very high uncertainties in the properties of this population, we prefer not to include it in our analysis. Therefore, we do not further discuss this type of binary in this work.

\subsection{Simulating GW catalogues}
We model the average intrinsic population of detectable compact objects of type $[s]$, with $s$ being BBH or BNS, according to 
\begin{equation}
    \frac{dN^{[s]}}{dVdtd\vec{\theta}}(\vec{\theta},z) = p^{[s]}\left(\vec{\theta}\right)\, R^{[s]}(z)~, 
    \label{eq:dN_dVdzdtddtheta}
\end{equation}
where $t$ is the time at the emitter and $\vec{\theta}$ the intrinsic parameters of the binary systems given by
\begin{equation}
    \vec{\theta} = \begin{pmatrix}
        m_1 & m_2 & \iota & \chi_1 & \chi_2  & \Lambda_1 & \Lambda_2 & \Phi_c
    \end{pmatrix}~.
\end{equation}
Here, $\iota$ is the inclination angle, $\Phi_c$ is the phase at coalescence, $m_i$ are the masses of the individual objects, $\chi_i$ are the spins,  and $\Lambda_i$ are the tidal deformabilities of the compact objects in the binary system, which are zero for BBH. 
In this prescription, we assume that the distribution $p^{[s]}(\vec{\theta})$ of the intrinsic parameters is independent of the redshift. Moreover, we imposed that the masses, spins and inclination angle are uncorrelated,
\begin{equation}
    p^{[s]}\left(\vec{\theta}\right) = p^{[s]}(m_1,m_2)\, p^{[s]}(\chi_1,\chi_2)\, p^{[s]}(\iota) \, p^{[s]}(\Lambda_1, \Lambda_2) p^{[s]}(\Phi_c)~.
\end{equation}
For BBH, we use the POWER LAW+PEAK distribution for the primary mass and a uniform distribution between $m_{\rm min}^{\rm BBH}=2.5\, M_\odot$ and $m_1$ for the secondary mass~\cite{{LIGOScientific:2020kqk,KAGRA:2021vkt}}. For the BNS, we adopt a uniform distribution between $1$ and $2.5 \, M_\odot$. Regarding the spins we take a Gaussian distribution with $\mu_\chi=0$ and $\sigma_\chi=0.1$ for BBH, while we neglect the impact of spins for BNS, following~\cite{Iacovelli:2022bbs}. We assume that the inclination angle is isotropically distributed for both populations. For BBH, the tidal deformabilities are zero, while for BNS we use a uniform distribution between 0 and 2000. For the phase at coalescence, we adopt a uniform distribution between $0$ and $2\pi$.

Following~\cite{Bellomo:2021mer}, we link the merger rate of BBH and BNS to the average star-formation rate (SFR) per halo evaluated by \texttt{UniverseMachine} \href{https://bitbucket.org/pbehroozi/universemachine/src/main/}{\faBitbucket} \cite{UniverseMachine}, by applying a time delay with $p(t_d)\sim 1/t_d$. This gives
\begin{equation}
    R^{[s]}(z) = \int dt_d \, p(t_d) \int dM_h \frac{dN_h}{dM_h}(z_f,M_h)\left\langle {\rm SFR}(z_f,M_h)\right\rangle~,
\end{equation}
with $dN_h/dM_h$ the halo mass function of~\cite{Tinker:2008ff}, while the formation redshift is defined as 
\begin{equation}
    z_f(t_d,z) \equiv z\left[t(z)-t_d\right]~.
\end{equation}
The merger rate is normalized at $z=0$ to the one measured by the LVK collaboration \cite{LIGOScientific:2020kqk,KAGRA:2021vkt,KAGRA:2021duu}, namely $R_{BBH}(z\!=\!0)=23.9$ Gpc$^{-3}$yr$^{-1}$ and $R_{BNS}(z\!=\!0)=105.5$ Gpc$^{-3}$yr$^{-1}$. Note in particular that $R_{BNS}$ still has a very large uncertainty, more than an order of magnitude. We conservatively take the best-fit value from \cite{KAGRA:2021duu}, but in principle a 10 times larger value is possible. 

As in~\cite{Begnoni:2024tvj}, the effect of the detector is modeled through the efficiency $\epsilon_{thr}$, which represents the fraction of CBC that is possible to resolve with the detector network considered. We evaluate it by selecting the events with signal-to-noise ratio (SNR) larger than a given threshold, $\rho \geq \rho_{\rm thr}$ in the networks HLVK, HLVIK and ET2CE. Specifically, we choose $\rho_{\rm thr}=12$ and compute the efficiency as
\begin{equation}
\epsilon_{thr}\left(\vec{\theta},z\right)=\frac{1}{4\pi}\int d\hat{n} \, \, \theta_{\rm Heaviside}\left[\rho\left(\vec{\theta},z,\hat{n}\right)-\rho_{\rm thr}\right]~.
\end{equation}
The SNR $\rho\left(\vec{\theta},z,\hat{n}\right)$ of the array of detector is calculated from the SNR of the single detectors SNR$_i$ as ${\rm SNR}^2 = \sum_i{\rm SNR}_i^2$ (see, e.g., Eq. (26) and (27) in \cite{Iacovelli:2022bbs}).
Finally, the total number of detected events per unit of redshift reads
\begin{equation}
    \frac{dN^{[s]}}{dz}(z) = \frac{4\pi c\ r^2(z)}{H(z)}\frac{T_{\rm obs}}{1+z}  \int d\vec{\theta} \, \,  \frac{dN^{[s]}}{dVdtd\vec{\theta}} (\vec{\theta},z) \  \epsilon_{thr}\left(\vec{\theta},z\right)~,
        \label{eq:dN_dz_W}
\end{equation}
with $r(z)$ the comoving distance at redshift $z$, $T_{\rm obs}$ the observing time,  $\frac{c\ r^2(z)}{H(z)}= \frac{dV}{dz}$, and the $1/(1+z)$ factor comes from the conversion from the source rest frame to the observer rest frame. We consider a total observing time of 10 years, $T_{\rm obs}=10$ yrs.

In practice, we evaluate the above integrals over $d\vec{\theta}$ and $d\hat{n}$ with a Monte Carlo method using the GWFAST code \cite{Iacovelli:2022bbs,Iacovelli:2022mbg}. A large number of GW events is sampled from the intrinsic distribution  \eqref{eq:dN_dVdzdtddtheta}. For each event, we evaluate the SNR and only retain the events with $\rho \geq \rho_{\rm thr}$. The redshift distribution of these events gives the un-normalized distribution $\frac{dN^{[s]}}{dz}(z)$ in \eqref{eq:dN_dz_W}. The normalization is fixed noting that, at $z=0$, the efficiency reaches 1, $\epsilon_{thr}\left(\vec{\theta},z=0\right)=1$, so that the integral in \eqref{eq:dN_dz_W} can be calculated directly. 

Finally, the  $\frac{dN^{[s]}}{dz}(z)$ distributions are calculated for each population (BBH and BNS) and detector configuration (HLVK, HLVIK, ET2CE) for our fiducial cosmology and converted into luminosity distance distributions
$\frac{dN^{[s]}}{dD_L}(D_L)$ (still using the same cosmology).

\begin{figure}[t]
    \centering
    \begin{subfigure}[b]{0.49\textwidth}
        \centering
        \includegraphics[width=\textwidth]{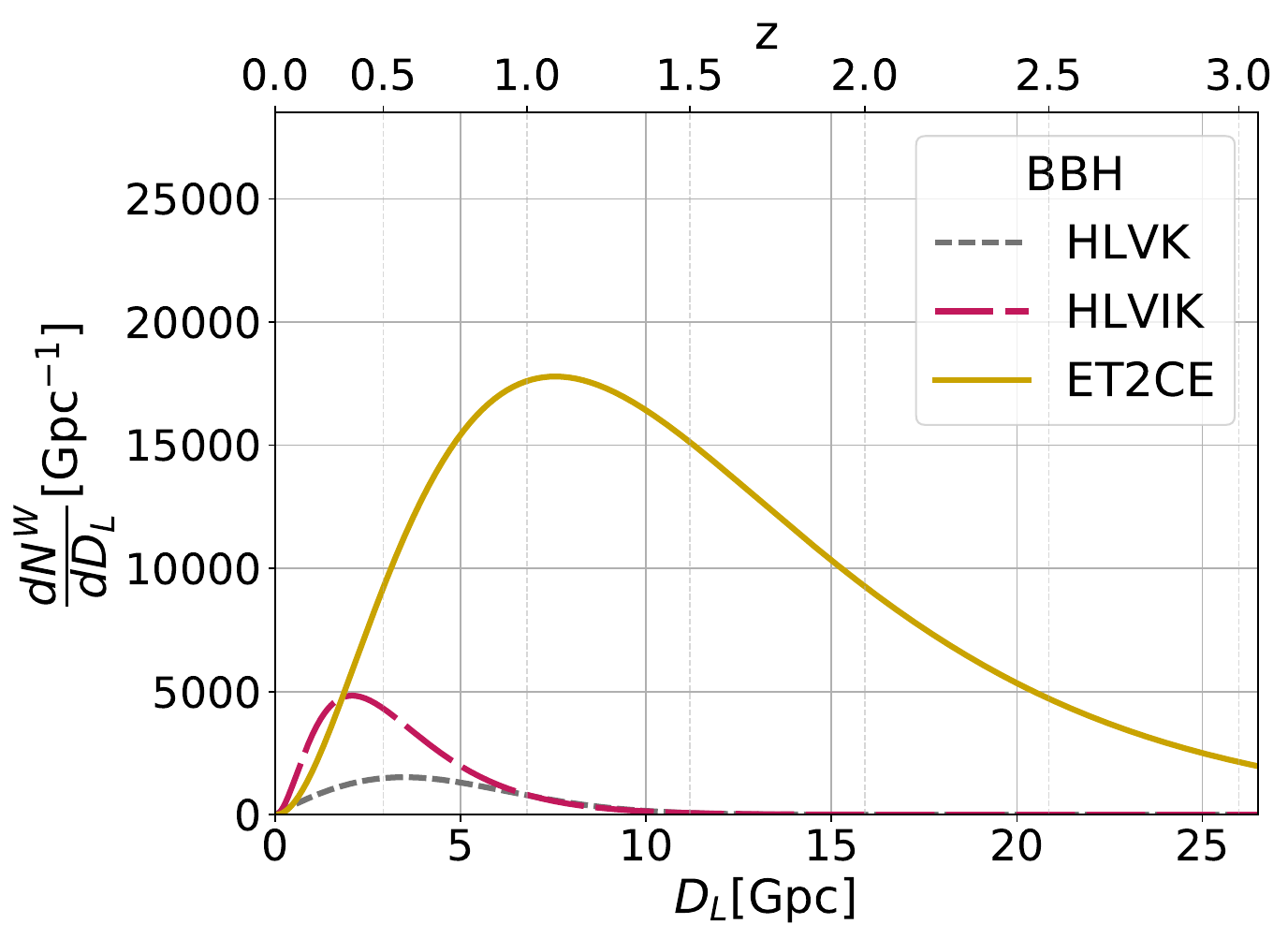}
        \caption{Detectable BBH  number density}
        \label{fig:BBH_distributions}
    \end{subfigure}
    \hfill
    \begin{subfigure}[b]{0.49\textwidth}
        \centering
        \includegraphics[width=\textwidth]{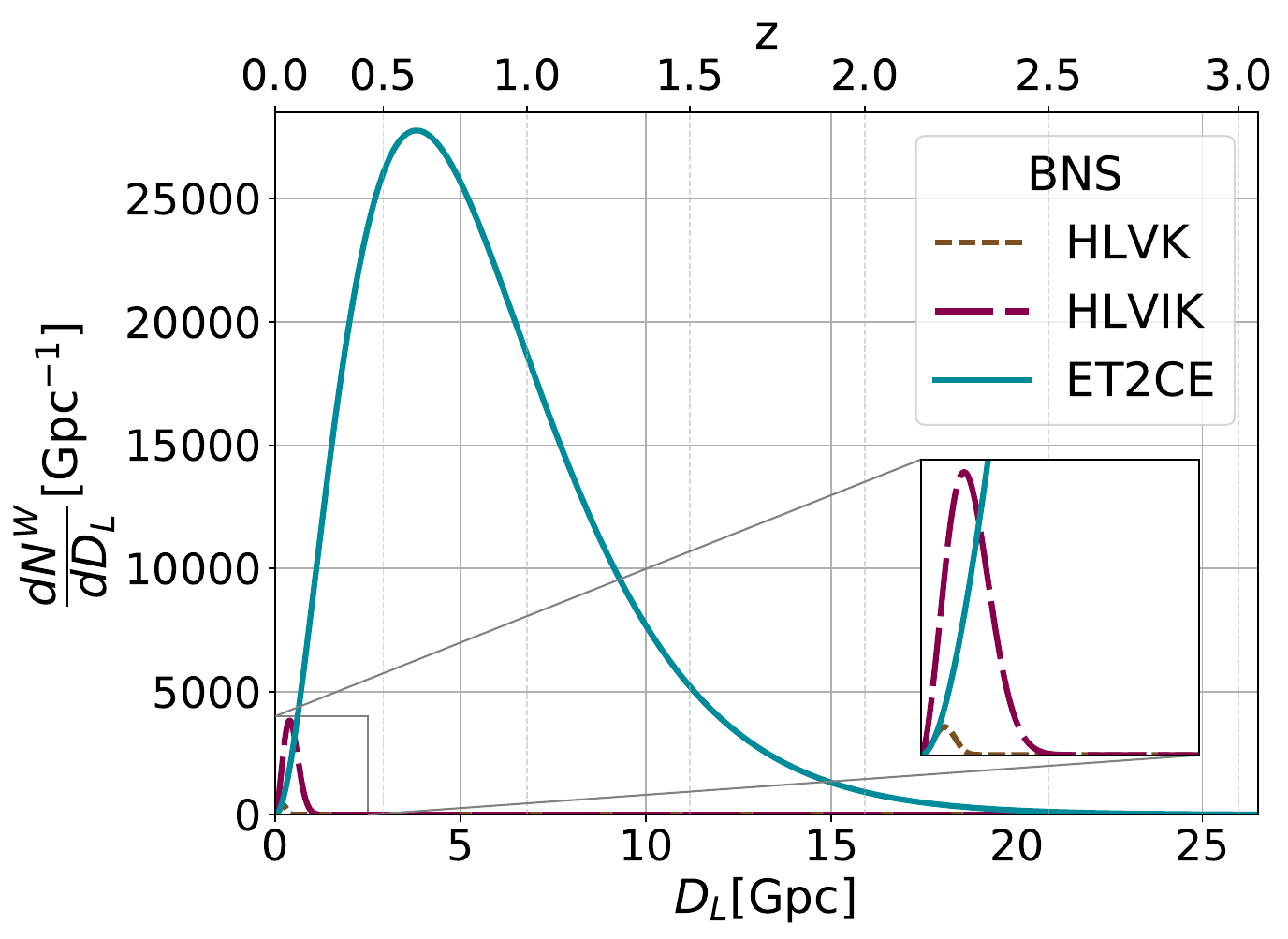}
        \caption{Detectable BNS  number density}
        \label{fig:BNS_distributions}
    \end{subfigure}

    \caption{Left: 
    Distribution of detectable BBH over distance for the 3 detector configuration considered. The upper $x$-axis shows the equivalent $z$ computed with the fiducial  cosmology.
    Right: the same, but for BNS.}
    \label{fig:BBH_BNS_distributions}
\end{figure}

\begin{table}[bp]
  \centering
  \begin{tabular}{|c||c|c|c|c|c|c|}
    \hline
    \textbf{Detector} & \textbf{Source} & \textbf{a} & \textbf{b} & \textbf{c} & \textbf{d} & \textbf{\# events} \\
    \hline
    \hline
    \multirow{2}{*}{\textbf{HLVK}} & BBH &  $1.24\cdot 10^{-3}$ & $9.31\cdot 10^{-1}$ & $1.97 \cdot 10^{-4}$ & $1.95$  & $9159$\\
    &BNS &  $1.22 \cdot 10^{-4}$ & $1.60$ & $3.86 \cdot 10^{-3}$ & $3.02$  & $86$ \\
    \hline
    \hline
    \multirow{2}{*}{\textbf{HLVIK}} & BBH &$3.04\cdot 10^{-6}$ & $2.23$ & $1.72 \cdot 10^{-3}$ & $8.04 \cdot 10^{-1}$ & $20289$\\
    &BNS &  $4.44\cdot 10^{-4}$ & $1.65$ & $2.30 \cdot 10^{-3}$ & $2.13$ & $1874$ \\
    \hline
    \hline
    \multirow{2}{*}{\textbf{ET2CE}} & BBH & $9.40\cdot 10^{-7}$ & $2.15$ & $3.76 \cdot 10^{-4}$ & $8.70 \cdot 10^{-1}$ & $262596$ \\
    &BNS & 
    $2.44\cdot 10^{-5}$ & $1.92$ & $4.97 \cdot 10^{-4}$ & $1.01$& $198257$\\
    \hline
  \end{tabular}
  \caption{Values of the fitting function~\eqref{eq:gen-fit} coefficients for the different detector configurations.}
  \label{tab:GW-distribution-numbers}
\end{table}

For ease of calculation, we fit to the above $\frac{dN^{[s]}}{dD_L}(D_L)$ an analytical form of the type \cite{Raccanelli:2016cud}: 
\begin{equation}
    \centering\label{eq:gen-fit}
    \frac{dN^{[s]}(D_L)}{dD_L} = a D_L^b \exp(-(c D_L)^d)~.
\end{equation}
The parameters $a$, $b$, $c$ and $d$ for the six different cases (2 populations x 3 detector configurations) are reported in Table \ref{tab:GW-distribution-numbers}.
The table also shows the number of events expected for each case in 10 years of data taking.
\autoref{fig:BBH_BNS_distributions} shows $\frac{dN^{[s]}}{dD_L}(D_L)$  for the 6 cases considered .
From the table and the plot it can be seen that only very few BNSs at low redshift will be detected by HLVK and HLVIK, not enough to have an impact on the analysis. For this reason, for the above two cases we will consider only BBHs. A significant number of BNSs is instead expected in the ET2CE case. In this case we will consider both BNS and BBH, and we will show the constraints achievable considering both of them separately and jointly.

\begin{figure}
    \centering
    \begin{subfigure}[b]{0.49\textwidth}
        \centering
    \includegraphics[height=0.8\textwidth]{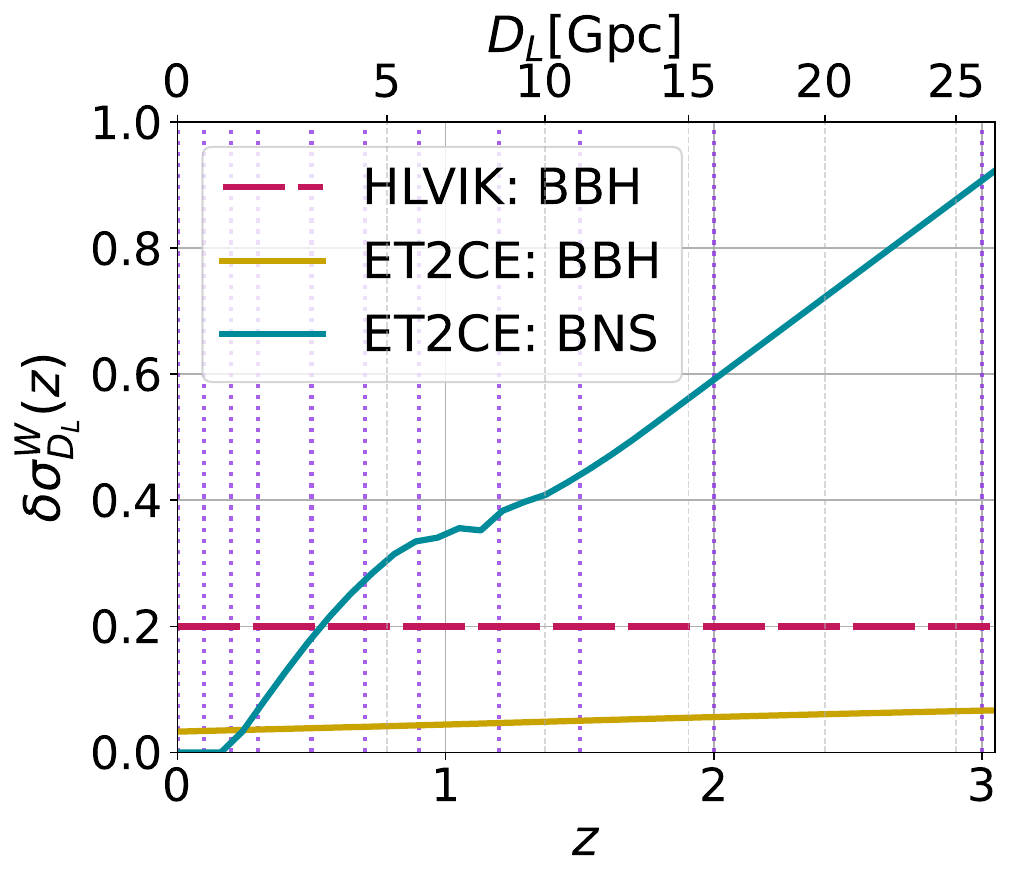}
    \caption{Luminosity distance relative error. }
    \label{fig:err_DL}
    \end{subfigure}
    \hfill
    \begin{subfigure}[b]{0.49\textwidth}
        \centering
    \includegraphics[height=0.8\textwidth]{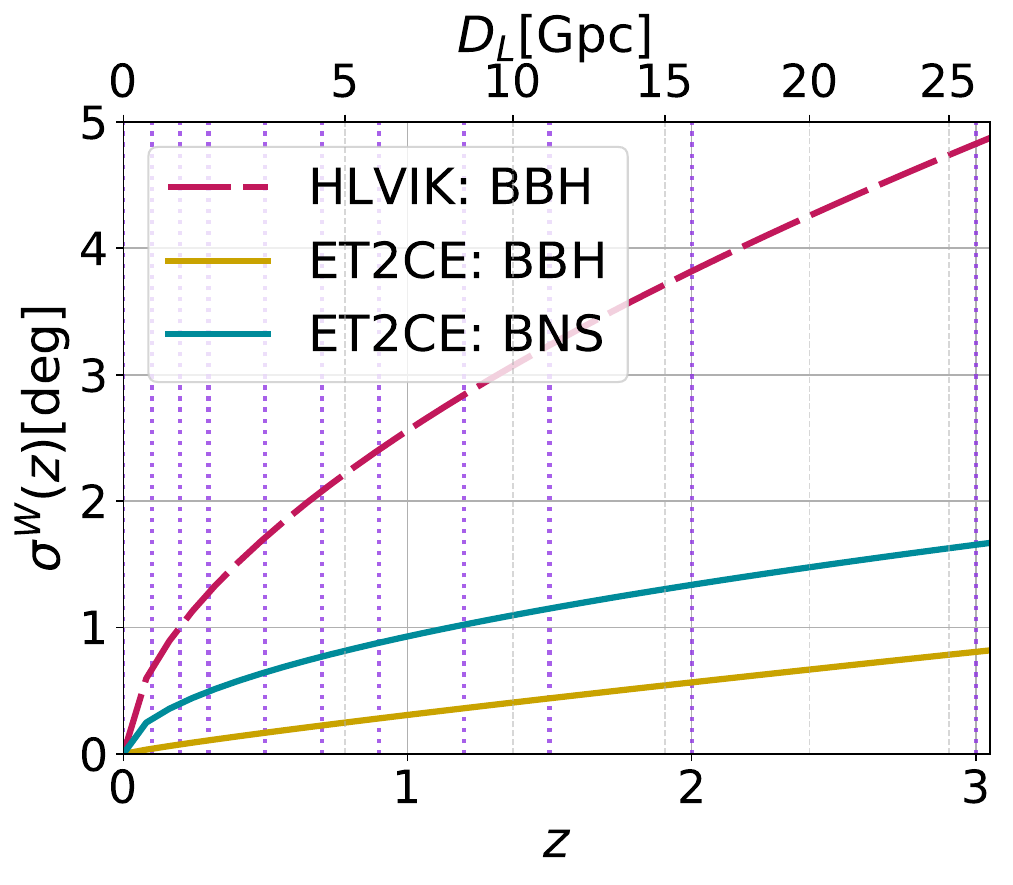}
    \caption{Angular localisation error.}
    \label{fig:err_sigma}
    \end{subfigure}
    \centering
    \caption{Left: Luminosity distance relative errors as a function of redshift employed in the analysis.  Right: Angular resolutions as a function of redshift employed in the analysis, as taken from \cite{Calore:2020bpd}.}
    \label{fig:err_DL_sigma}
\end{figure}

For a full characterization of the GW population seen by a given detector, besides $\frac{dN^{[s]}}{dD_L}(D_L)$, we need to know the angular resolution $\sigma^{[s]}$ and the luminosity distance error $\delta \sigma_{D_L}$, as outlined in the previous section.
Regarding the angular localisation error, we use the results derived in \cite{Calore:2020bpd}. In \cite{Calore:2020bpd} GW detector configurations equivalent to our HLVIK and ET2CE are analysed, and a specific study of the 
angular resolution for BBHs and BNSs
is derived. 
Specifically, a population of BBH and BNS events is simulated and the localization sky-map is reconstructed using the Bayestar software \cite{Singer:2015ema}. The localization is typically found to be approximately Gaussian and the $\sigma^{[s]}$ of the Gaussian is derived for each event. A $\sigma^{[s]}$ averaged over the population is then build, and this average angular resolution  is found to be redshift dependent $\sigma^{[s]}(z)$. In \autoref{fig:err_DL_sigma}, we plot $\sigma^{[s]}(z)$ for BBH and BNS and for  HLVIK and ET2CE as taken from \cite{Calore:2020bpd},  which will be used in our analysis. 
For the HLVK case,
which is not treated in \cite{Calore:2020bpd}, we use the same $\sigma^{[s]}(z)$ of HLVIK, although this corresponds to a slightly optimistic choice since with one detector less, the angular resolution will be slightly worse. 
Further studies of the angular resolution and sky localization of a network of GW events can be found in \cite{Wen:2010cr, Zhao:2017cbb}. When the network is composed of 3 or more detectors, as the cases discussed in the present analysis, the localization is typically well constrained without multimodality. Deviation from Gaussianity can instead appear, but are here neglected for the sake of simplicity, with a more detailed study of this effect left for  future works.

Following~\cite{Begnoni:2024tvj}, to derive the relative luminosity distance error $\delta \sigma_{D_L}$, we use the events generated through the GWFAST simulation.  For each simulated event, the code calculates the error on the relevant event quantities, including $\delta \sigma_{D_L}$, via a Fisher matrix formalism. In this way, the average population error can be derived as a function of $z$, see  \autoref{fig:err_DL_sigma}. 
It can be seen that for BBH $\delta \sigma_{D_L}$ is roughly constant in $z$ and equal to 5\% in the ET2CE case, and 20\% in the HLVIK case.
For BNS and ET2CE $\delta \sigma_{D_L}$ is worse and reaches $40\%$ at redshift $z\sim 1$.
In principle, with the same methodology, $\sigma^{[s]}(z)$ can also be derived.  We verified that the GWFAST-derived $\sigma^{[s]}(z)$ and the ones from \cite{Calore:2020bpd} that we actually use are in agreement at about 20\%, with residual differences that we attribute mainly to different parameters for the parent BBH and BNS populations from which the GW events are simulated.

\begin{figure}
    \centering 
    \begin{subfigure}[b]{0.49\textwidth}
        \centering
    \includegraphics[width=\textwidth]{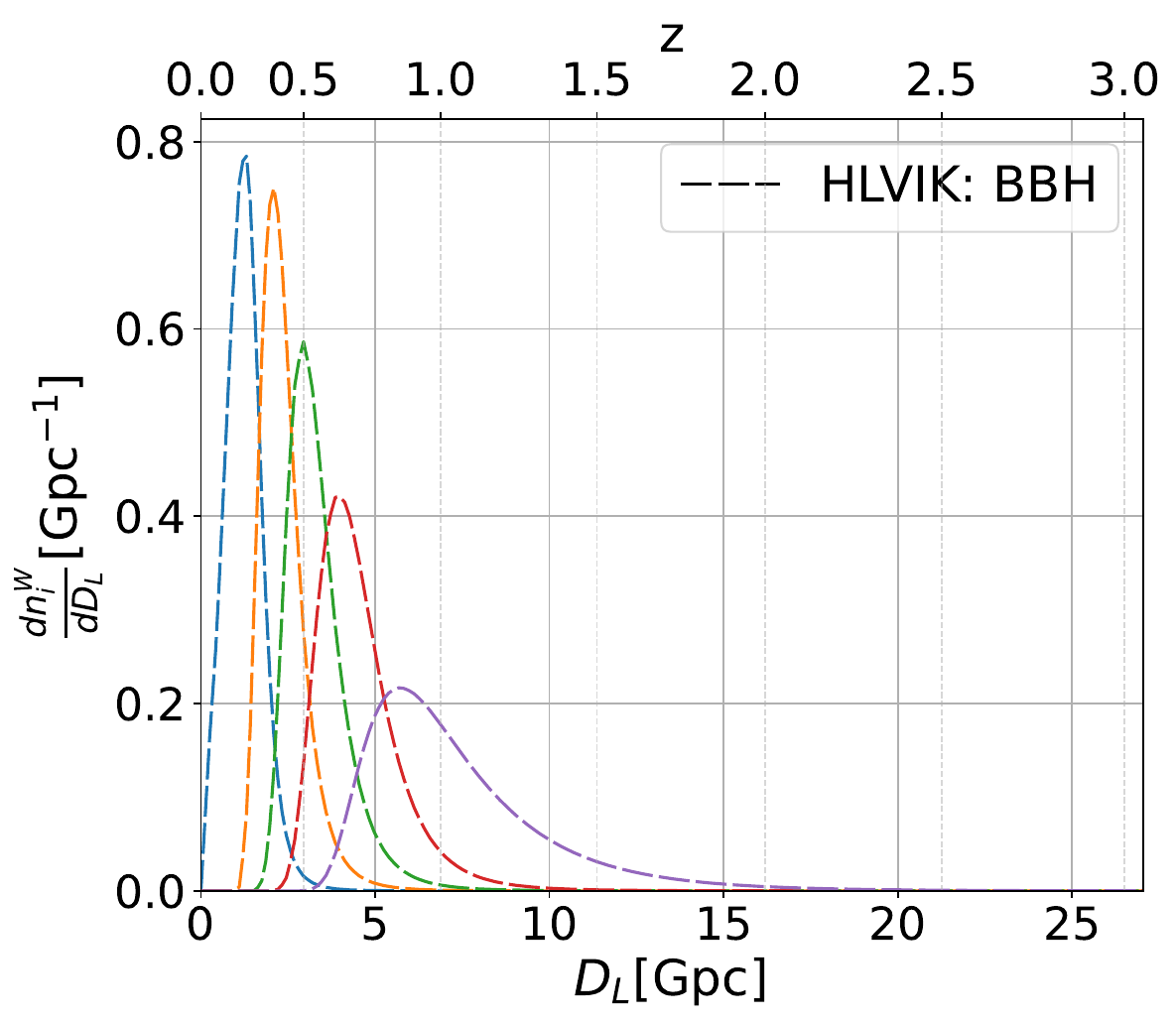}
    \caption{HLVIK binned distribution}
    \label{fig:bin_dist_HLVIK}
    \end{subfigure}
    \hfill
    \begin{subfigure}[b]{0.49\textwidth}
        \centering
    \includegraphics[width=\textwidth]{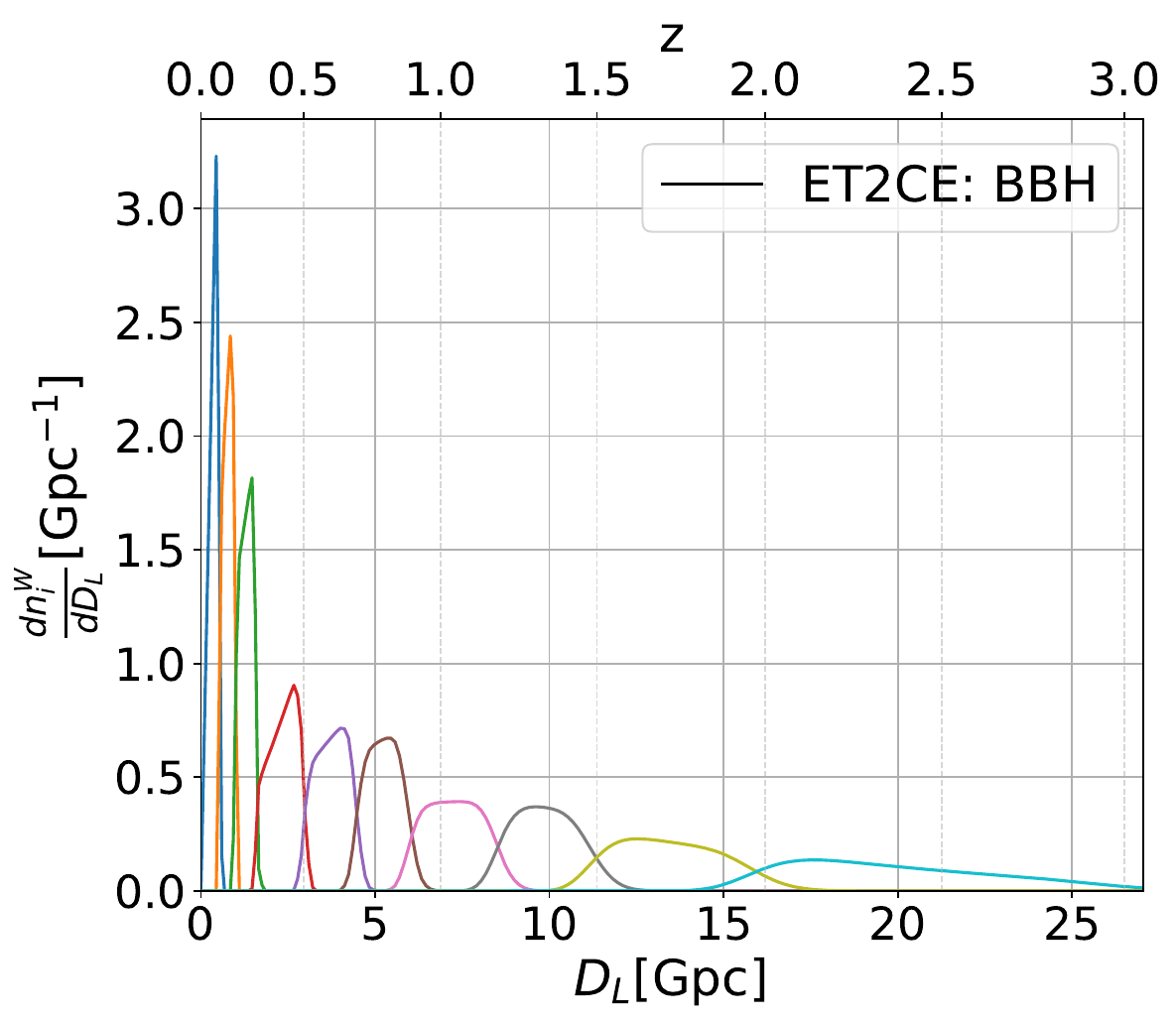}
    \caption{ET2CE binned distribution}
    \label{fig:bin_dist_ET2CE}
    \end{subfigure}
    \centering
    \caption{Left:  normalised density distribution $dn^{W_i}/dD_L$ as a function of $D_L$ for each of our 5 $D_L$-bins for the HLVIK BBH case.
    The upper $x$-axis shows redshifts converted from $D_L$ using the fiducial cosmology. Right: same for the ET2CE BBH case.}
    \label{fig:bin_dist_W}
\end{figure}

Given the above ingredients, similarly to what was done for the galaxies, we can choose a $D_L$-binning and derive the distribution $\frac{dn(D_L)}{dD_L}$ of \eqref{eq:conv_n_i_W} in each bin, i.e., the normalised distributions of GW events taking into account the error in luminosity distance.
In each bin we assume a constant  $\sigma_i^{[s]} = \sigma^{[s]}(z_i)$ and $\delta \sigma_{i\, D_L} = \delta \sigma_{D_L}(z_i)$   equal to the value at the center of the bin. 
For the ET2CE case, we choose 10 bins corresponding to the same 10 $z$-bins used for the galaxy catalogue, but converted to $D_L$ using our fiducial cosmology. For BBHs, all 10 bins are used, while for BNSs only the first 9 are relevant due to the fact that the last bin is empty (see \autoref{fig:BBH_BNS_distributions}).
For HLVK and HLVIK, instead, given the overall low number of events and average lower redshift with respect to ET2CE, we choose to have 5 bins with a maximal $D_L$ correspondent to $z=2.0$ and the other edges chosen in such a way to have an equal number of events in each bin.
 In this way, we still have a reasonable number of events in each bin to minimise the Poisson noise, and the number of bins is still sufficiently large to provide tomographic information. The edges of the luminosity distance bins are reported in \autoref{tab:DLbins}, while
\autoref{fig:bin_dist_W}  shows the BBH $\frac{dn(D_L)}{dD_L}$ for each bin for the HLVIK and ET2CE cases. 

\begin{table}[h!]
  \centering
  \begin{tabular}{|c||c|l|}
    \hline
    \textbf{Detector} & \textbf{Source} & \textbf{$D_L^{\rm edges}$~ [${\rm Mpc}$]} \\\hline
    \hline
    \textbf{HLVK}&BBH&$[0, 2326, 3568, 4824, 6450, 15905]$
    \\\hline\hline
    \textbf{HLVIK}& BBH&$[0, 1650, 2499, 3461, 4887, 15905]$
    \\\hline\hline
    \multirow{2}{*}{\textbf{ET2CE}}
    & BNS & $[0, 475, 1010, 1599, 2914, 4377, 5956, 8490, 11178, 15905]$\\
    &BBH
    & $[0, 475, 1010, 1599, 2914, 4377, 5956, 8490, 11178, 15905, 25987]$\\
    \hline
  \end{tabular}
  \caption{Luminosity distance bin edges of the dark sirens distribution for the various detector configurations and sources. 
  }
  \label{tab:DLbins}
\end{table}

\section{Likelihood Calculation}\label{sec:likelihood}

The likelihood formalism for cosmology inference using galaxy catalogues is well established. The likelihood functional form can be easily derived starting from a Gaussian likelihood for the $a_{\ell m}$ coefficients of the map and averaging over $m$ assuming statistical isotropy (see, e.g, \cite{euclidcollaboration2024euclidpreparation6x2pt,Casas:2022vik}). 
In the present analysis, we will use 
an approximate form of the likelihood in which the $C_\ell$s, instead of the $a_{\ell m}$s, are assumed to be Gaussian-distributed,
which we dub the $C_\ell$-based likelihood. Nonetheless, we have also implemented the full $a_{\ell m}$-based likelihood and compared the posteriors derived from the two formalisms in a few relevant cases without finding appreciable differences.
\autoref{fig:alm_bin_Triangle} in \autoref{app:triangle} shows the agreement between the $a_{\ell m}$-based and $C_\ell$-based likelihood results.
The main advantage of the $C_\ell$-based likelihood is its 
flexibility and ease of interpretation of the various terms. In particular, as we will see, it is straightforward to isolate and consider the different terms separately, as the ones related to the pure galaxy-GW cross-correlation.

Given the above considerations, the explicit form of the $C_\ell$-based likelihood is:
\begin{equation}
        \centering\label{eq:cl_likelihood}
         -2\, \text{ln}\mathcal{L}(\vec{D}|\vec{\theta}) = \sum_{\ell=\ell_{\rm min}}^{\ell_{\rm max}} \left(\vec{D}_\ell-\vec{T}_\ell(\vec{\theta})\right)^T \mathcal{C}^{-1}_\ell \left(\vec{D}_\ell-\vec{T}_\ell(\vec{\theta})\right)~,
\end{equation}
where $\theta$ represents a set of cosmological parameters that need to be inferred, including the bias coefficients from \eqref{eq:bias_gc} and \eqref{eq:bias_gw}, $\vec{D}_\ell$ is our data vector for multipole $\ell$, $\vec{T}_\ell$ is the theory vector for the same multipole, which depends on the free parameters $\vec{\theta}$, and finally $\mathcal{C}_\ell$ is the covariance matrix for this multipole. We use $\ell_{\rm min}=2$ and $\ell_{\rm max}=1000$\footnote{
We adopt the Limber approximation throughout our work \cite{LimberApprox,LoVerde2008}, which is shown in \cite{LoVerde2008} to be precise at about a 10\% level for autocorrelations and 1\% for cross-correlations at relatively low $\ell\leq 100$. 
At high $\ell \geq 200-300$ instead non-linear corrections to the matter power spectrum become important, which we take into account using the \texttt{halofit} non-linear correction in  \texttt{CLASS} \cite{Takahashi2012}.}.

Both $\vec{D}_\ell$ and $\vec{T}_\ell$ have a similar structure, which can be summarised as:
\begin{align}\centering\label{eq:vectors}
    \vec{V}_\ell&=
    \begin{pmatrix}
 \overrightarrow{GG}_\ell\\
 \overrightarrow{WG}_\ell\\
 \overrightarrow{WW}_\ell\\
\end{pmatrix}.
\end{align}
Here, $\overrightarrow{GG}_\ell= C_{j\ge i}^{G_iG_j}(\ell)$ is the sub-vector containing all the galaxy autocorrelations and cross-correlations among the different redshift bins given in \eqref{eq:Cl_GG}. Since we have $n^G=10$ bins, the vector contains $[n^G \times (n^G+1)]/2 = 55$ entries for a given $\ell$.
Similarly, we have $\overrightarrow{WW}_\ell= C_{j\ge i}^{W_iW_j}(\ell)$ for gravitational waves, with $[n^W \times (n^W+1)]/2$ entries. This time we use either $n^W=5$ or 10, depending on the detector configuration.
Finally, $\overrightarrow{WG}_\ell= C^{W_iG_j}(\ell)$ contains $n^G n^W$ entries.  The vector $\vec{V}(\ell)$ has thus a total of $[(n^G+n^W) \times  (n^G+n^W+1)]/2$ entries for each $\ell$.

Since we are performing a forecast, the data vector is just the theory vector evaluated at a fiducial cosmology,
\begin{align}
    \centering\label{eq:datatheory_vector}
    \vec{D}_\ell&=\vec{T}_\ell(\vec{\theta}_{\rm fid})~, 
\end{align}
where $\vec{\theta}_{\rm fid}$ is the set of fiducial parameters. Assuming a Gaussian distribution of the spherical harmonic coefficients, the covariance matrix $\mathcal{C}_\ell$ reads \cite{euclidcollaboration2024euclidpreparation6x2pt,Casas:2022vik}:
\begin{align}
{\rm Cov}\left[ \vec{T}_\ell(\vec{\theta}), \vec{T}_{\ell'}(\vec{\theta}) \right]
&= {\rm Cov}\left[C^{A_iB_j}(\ell),C^{C_kD_n}(\ell')\right]
\nonumber \\
&=  \frac{\delta_{\ell\ell'}}{(2\ell+1)f_{\rm fov}\Delta\ell}  \Big[ \left( C^{A_iC_k}(\ell) + N^{A_iC_k}(\ell) \right)   \left( C^{B_jD_n}(\ell) + N^{B_jD_n}(\ell) \right) + \nonumber\\
&  + \left( C^{A_iD_n}(\ell) + N^{A_iD_n}(\ell) \right) \left( C^{B_jC_k}(\ell) + N^{B_jC_k}(\ell) \right)  \Big]~, 
\label{eq:covmat}
\end{align}
where $A,B,C,D$ can either be $G$ or $W$, and $f_{\rm fov}$ is the fraction of sky where the galaxy survey and GW survey overlap. Since GW experiments are full-sky experiments, $f_{\rm fov}$ coincides with $f_{\rm fov}^G$. 
Note that the covariance matrix depends on the noise as defined in \eqref{eq:noise_def}.
This analysis can be sped up by being performed on a binned $\ell$-space, instead of being evaluated for each multipole.
$\Delta\ell$ represents the width of the bin in $\ell$-space.
In order to achieve optimal accuracy, our main results have been computed with $\Delta\ell=1$.
However, we have tested that using $\Delta\ell=20$ would not significantly alter the results. The explicit comparison of the two cases is shown in \autoref{fig:alm_bin_Triangle} of Appendix~\ref{app:triangle}.
For a given $\ell$, the above square matrix has $[(n^G+n^W)^2 \times (n^G+n^W+1)^2]/4$ elements
and can be written in the reduced form 
\begin{equation}\centering\label{eq:total_covmat_red}
\mathcal{C}_\ell=
    \begin{pmatrix}
        (\overrightarrow{GG}_\ell,\overrightarrow{GG}_\ell) & (\overrightarrow{GG}_\ell,\overrightarrow{WG}_\ell) & (\overrightarrow{GG}_\ell,\overrightarrow{WW}_\ell)\\
        (\overrightarrow{WG}_\ell,\overrightarrow{GG}_\ell) & (\overrightarrow{WG}_\ell,\overrightarrow{WG}_\ell) & (\overrightarrow{WG}_\ell,\overrightarrow{WW}_\ell)\\
        (\overrightarrow{WW}_\ell,\overrightarrow{GG}_\ell) & (\overrightarrow{WW}_\ell,\overrightarrow{WG}_\ell) & (\overrightarrow{WW}_\ell,\overrightarrow{WW}_\ell)\\
    \end{pmatrix}~.
\end{equation}
In order to calculate the full likelihood of \eqref{eq:cl_likelihood}, this matrix needs to be inverted for each separate $\ell$. We can also consider separately the information coming from galaxy catalogues alone, from gravitational waves alone, or only from the cross-correlation between the two probes. In these cases, for each $\ell$, the data vector reduces, respectively, to $\overrightarrow{GG}_\ell$, $\overrightarrow{WG}_\ell$ or $\overrightarrow{WW}_\ell$, while the covariance matrix is given by one of the diagonal blocks $(\overrightarrow{GG}_\ell,\overrightarrow{GG}_\ell)$, $(\overrightarrow{WW}_\ell,\overrightarrow{WW}_\ell)$ or $(\overrightarrow{WG}_\ell,\overrightarrow{WG}_\ell)$. 

We use the above likelihood to sample the space of model parameters $\vec{\theta}$, using the Metropolis-Hastings MCMC algorithm as implemented in \texttt{MontePython}. We sample over the five cosmological parameters 
$\{ \omega_{\rm b}, \omega_{\rm cdm}, n_{\rm s}, A_{\rm s}, h\}$ of the $\Lambda$CDM model,
plus the three nuisance bias parameters defined in Eqs.~\eqref{eq:bias_gc} and \eqref{eq:bias_gw}, $\{a_1^G, a_1^W, a_2^W\}$. \autoref{tab:cosmo_nuisance} shows our choice of fiducial values and flat prior edges for each parameter.

\begin{table}[t]
    \centering
    \begin{tabular}{|c|c|c|c|}
        \hline
        \textbf{Parameter} & \textbf{Description} & \textbf{Fiducial value} & \textbf{Prior} \\
        \hline
        $\omega_{\rm b}$ & baryon density & $2.249 \times 10^{-2}$ & $[10^{-4},1]$ \\
        $\omega_{\rm cdm}$ & cold dark matter density & $0.112$ & $[10^{-2},0.5]$ \\
        $n_{\rm s}$ & scalar spectral index & $0.96605$ & $[0.8, 1.2]$ \\
        $A_{\rm s}$ & scalar amplitude & $2.42 \times 10^{-9}$ & $[0.0,10^{-7}]$ \\
        $h$ & hubble parameter & $0.6737$ & $[0.1, 1.5]$ \\
        \hline
        $a_1^G$ & galaxy clustering bias normalization & $1.0$ & $[0.5, 2.0]$ \\
        $a_1^W$ & GW bias normalization & $2.0$ & $[0.0, 4.0]$ \\
        $a_2^W$ & GW bias slope & $0.0$ & $[-2.0, 7.0]$ \\
        \hline
    \end{tabular}
    \caption{Fiducial values and flat prior edges for the cosmological and nuisance parameters used in our forecast.}
    \label{tab:cosmo_nuisance}
\end{table}

\begin{figure}[t]
    \centering
        \includegraphics[width=0.7\textwidth]{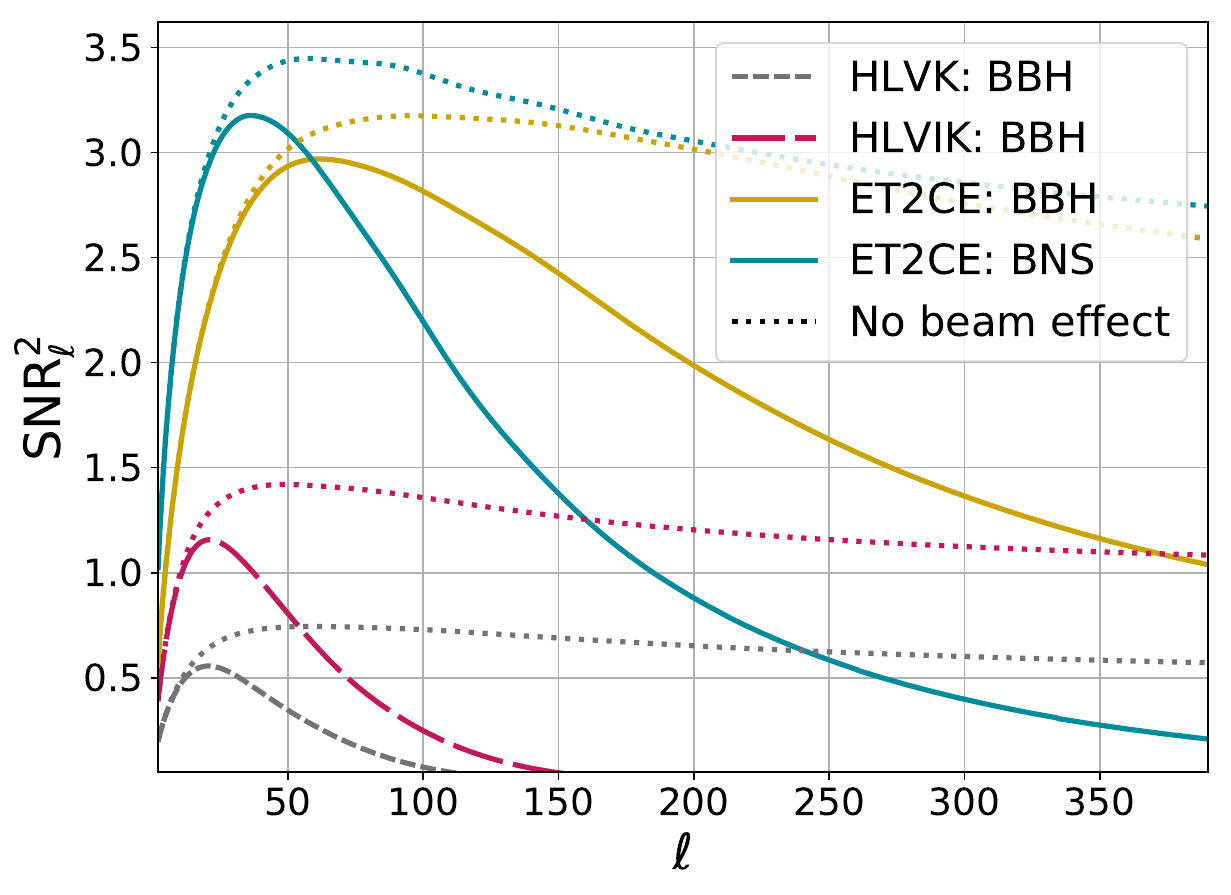}
    \caption{
    Cross-correlation SNR contributions as function of $\ell$ (see  \eqref{eq:SNR2}) for the various GW source and detector configurations considered in the analysis for 10 years of observation. The solid line shows the actual SNR, while the dotted line shows the results assuming perfect localisation, i.e, no beam effect, $\sigma^W_i=0$ for all bins in \eqref{eq:beam_ang}, illustrating the degradation in sensitivity due the finite angular resolution}. The total SNR is reported in \autoref{tab:SNR}. 
        \label{fig:SNR}
\end{figure}

\begin{table}[ht]
  \centering
  \begin{tabular}{|c||c|c|}
    \hline
    \textbf{Detector} & \textbf{Source} & \textbf{SNR} \\
    \hline
    \hline
    \textbf{HLVK}
    & BBH & $3.6$\\
    \hline
    \hline
\textbf{HLVIK}
    & BBH &  $7.9$ \\
    \hline
    \hline
    \multirow{2}{*}{\textbf{ET2CE}}    
    & BBH &  $44.0$ \\
    & BNS &  $31.9$ \\
    \hline
  \end{tabular}
  \caption{Total SNR for the various cases shown in \autoref{fig:SNR}.  }
  \label{tab:SNR}
\end{table}

The likelihood can also be used to assess the overall sensitivity of the probe through the signal-to-noise-ratio SNR \cite{Scelfo_2020}. In practice, the expected SNR can be evaluated by simply taking $\vec{D}_\ell=0$ and $
\vec{\theta}=\vec{\theta}_{\rm fid}$ in the likelihood, i.e.,
\begin{equation} \label{eq:SNR2}
{\rm SNR}^2 = \sum_{\ell=\ell_{\rm min}}^{\ell_{\rm max}}{\rm SNR}_\ell^2=\sum_{\ell=\ell_{\rm min}}^{\ell_{\rm max}}\vec{T}_\ell(\vec{\theta_{\rm fid}})^T \ \mathcal{C}^{-1}_\ell \ \vec{T}_\ell(\vec{\theta_{\rm fid}})~,
\end{equation}
and we only consider the contributions coming from the GW auto-correlation, $\overrightarrow{WW}_\ell$, and galaxy-GW cross-correlation, $\overrightarrow{WG}_\ell$, i.e., we exclude the galaxy auto-correlations, $\overrightarrow{GG}_\ell$.
We show in \autoref{fig:SNR} the contribution to the SNR$^2$ coming from each term in sum in \eqref{eq:SNR2} as function of multipole $\ell$ for the various GW sources and detector configuration considered. We report the total SNR summed over all the $\ell$s in \autoref{tab:SNR}. We find that the SNR expected in the HLVK and HLVIK cases are quite low, peaking around 1 and giving a total SNR of 3.6 and 7.9 respectively. Thus, cosmological constraints in these two cases are expected to be loose. Indeed, we will see that this expectation will be confirmed in the next section. The ET2CE case has instead a much larger SNR, and it is thus expected to have a significant constraining power.  The plot also shows the SNR in the ideal case of infinite GW angular resolution, showing that the finite angular resolution has a critical impact on this kind of analysis.

\section{Results}\label{sec:results}

In this section, we report the results of our MCMC likelihood scans for the various GW detectors considered. 
As explained in the previous sections, for the HLVK and HLVIK configurations, we only consider the contribution from the BBH population, while for the ET2CE configuration, we consider both BBHs and BNSs. In particular, we analyse the three cases of BBH-only, BNS-only, and BBH and BNS together.

For each GW configuration, we further consider four sub-cases, depending on which part of the covariance matrix we include in the likelihood: galaxy auto-correlation only (GG), GW auto-correlation only (WW), GW$\times$galaxy cross-correlation only (XC), and full covariance matrix (GG WW XC). Note that, for better clarity, we indicate the cross-correlation likelihood as XC in order to reserve the notation WG for the cross-correlation spectra $C_\ell^{WG}$.

\begin{table}[t]
  \centering
  \begin{tabular}{|c||c|c|}
    \hline
    \textbf{Detector} & \textbf{Contribution} & \makecell{\textbf{$H_0$ bestfit$_{-1\sigma}^{+1\sigma}$} \\ 
    $\left[{\rm km}\,{\rm s}^{-1}\,{\rm Mpc}^{-1}\right]$} \\
    \hline
    \hline
    \makecell{\textbf{Galaxy}\\\textbf{Catalogue}}
    & GG & $67.4_{-2.8}^{+2.3}$ \\
    \hline
    \hline
    \multirow{2}{*}{\textbf{HLVK}} 
    & XC & $67.3_{-24.2}^{+15.2}$\\
    & Full matrix &  $67.6_{-2.9}^{+2.2}$\\
    \hline
    \hline
    \multirow{2}{*}{\textbf{HLVIK}}    
    & XC & $67.3_{-5.2}^{+4.5}$ \\
    & Full matrix & $67.5_{-2.5}^{+2.0}$ \\
    \hline
    \hline
    \multirow{2}{*}{\textbf{ET2CE}}    
    & XC & $67.3_{-0.9}^{+0.8}$\\
    & Full matrix & $67.2_{-0.4}^{+0.5}$ \\
    \hline
  \end{tabular}
  \caption{Hubble constant $H_0$ best-fit and error for the various cases considered in the text.  The fiducial value of $H_0$ is 67.4.} 
  \label{tab:h_results}
\end{table}

\begin{figure}[ht]
    \hspace{0.1\textwidth}
    \includegraphics[height=0.34\textwidth]{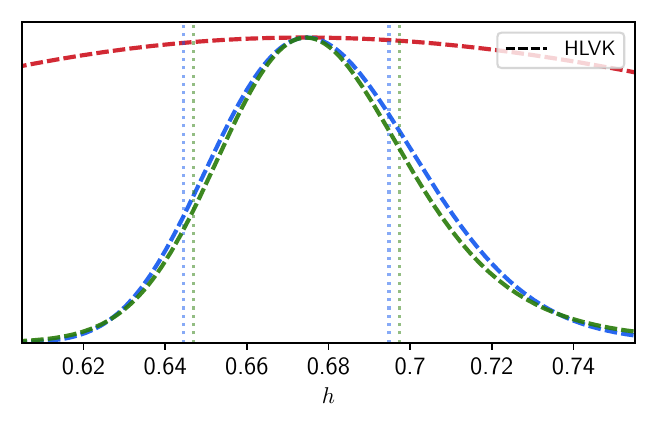}
    
    \hspace{0.1\textwidth}
    \includegraphics[height=0.34\textwidth]{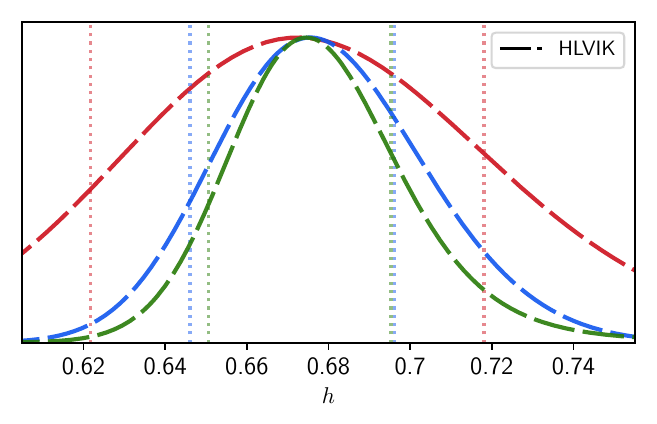}
    \includegraphics[height=0.34\textwidth]{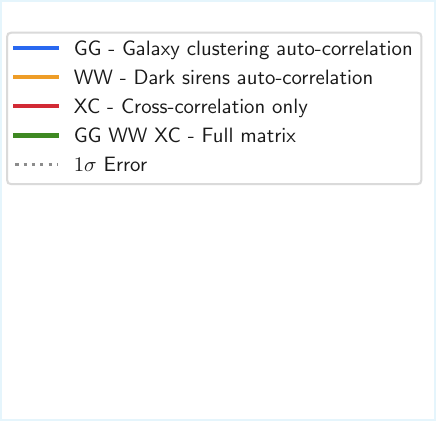}
    
    \hspace{0.10\textwidth}
    \includegraphics[height=0.34\textwidth]{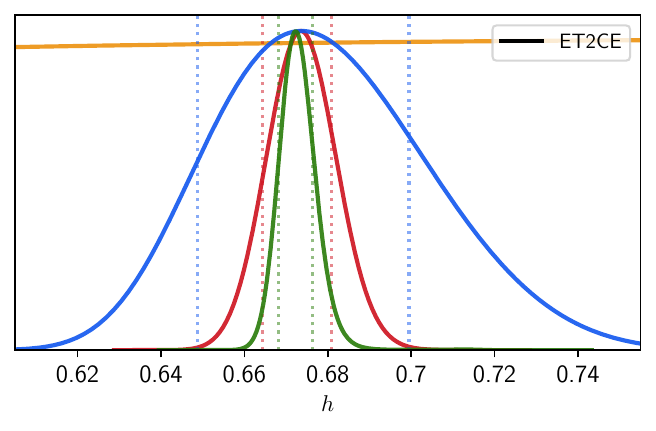}
    \caption{Posterior on the Hubble constant $h$ for the various cases considered in the analysis. From top to bottom the three detector configurations HLVK, HLVIK, ET2CE are shown, and for each of them the sub cases GG, XC, WW and full matrix. The dotted vertical lines show the $1\sigma$ intervals.}
    \label{fig:components}
\end{figure}

\begin{figure}[ht]
    \includegraphics[width=\textwidth]{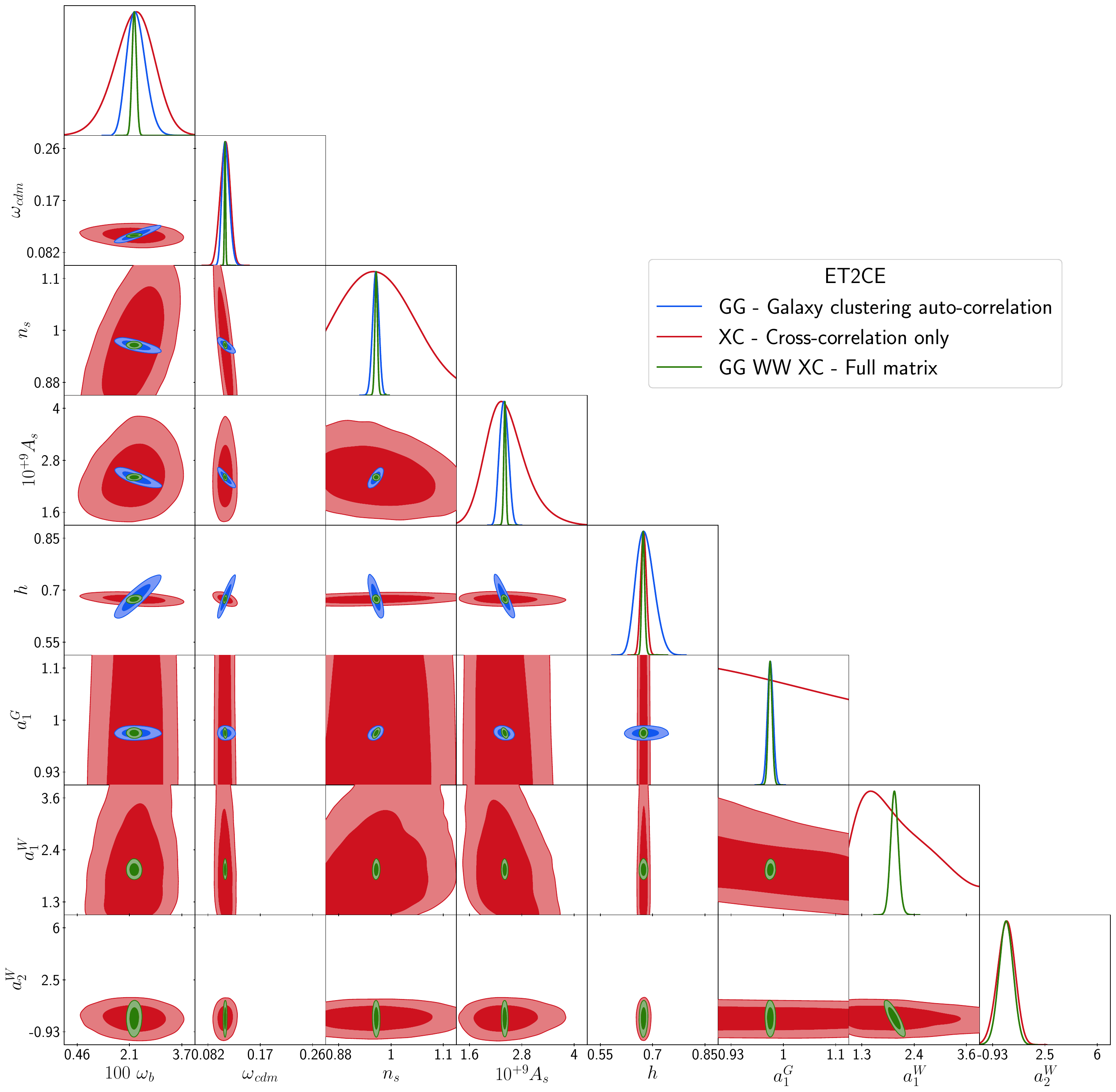}
    \caption{Full triangle plot of the MCMC scans for the GG, XC, and full matrix cases considering both BBH and BNS detections with ET2CE. $1\sigma$ and $2\sigma$ contours are shown. Information is not shown for parameters that are not present in a specific run, e.g. $a_1^W$ and $a_2^W$  for the GG scan.
    The numerical constraints of this study case are reported in \autoref{tab:results_ET_2CE}, in \autoref{app:numresults}.} 
    \label{fig:ET_2CE_Triangle}
\end{figure}

\subsection{Galaxy Catalogue Constraints}

We show in \autoref{fig:components} and \autoref{tab:h_results} the constraints on the Hubble parameter $h$ and Hubble constant $H_0$ obtained when only the information from the galaxy catalogue is used (case dubbed GG). We get a relative error of about 4\%, which will be our reference sensitivity when discussing the constraining power of the WW and XC probes.\footnote{Our result is in line with the official forecasts for Euclid \cite{Euclid:2019clj,EUCLID:2023uep,Euclid:2024imf} or LSST \cite{Zhang:2025rxp}, although a direct comparison is difficult since there the information from galaxy photometric clustering is always considered in combination with weak lensing.}
We stress that better constraints are achievable using also the spectroscopic samples and  lensing information \cite{Euclid:2019clj,EUCLID:2023uep,Euclid:2024imf}, but here
for consistency we only compare the constraining power of the cross-correlation between the GWs and photometric galaxy samples with that of the auto-correlation of the same photometric galaxy sample.

\subsection{HLVK and HLVIK Constraints}
For the HLVK configuration, we can see from \autoref{fig:components} and \autoref{tab:h_results} that the XC case only has a very loose sensitivity to $h$ of the order of 30\%. 
The situation improves significantly for the HLVIK XC case, which provides a 7\% sensitivity, indicating that the addition of a single GW detector, in this case LIGO India, dramatically improves the localisation capabilities of the detectors and thus the power of the cross-correlation analysis.
Interestingly, in this case,  XC  has a similar sensitivity to  GG, and the joint constraints provide a slight improvement with respect to the GG-only case.
Note that the XC case for the HLVK and HLVIK configurations has no sensitivity to $\omega_{\rm b}$, $n_{\rm s}$ and $A_{\rm s}$, and thus we fixed these parameters in the corresponding MCMC runs.
We do not report the WW results for these configurations since they  did not provide any constraint. 
The full triangle plots for all our HLVK and HLVIK forecasts are reported for completeness in \autoref{fig:LVK_Triangle} and \autoref{fig:HLVIK_Triangle} of Appendix~\ref{app:triangle}.

Our results are in agreement with \cite{Ferri_2025}, which finds an achievable sensitivity on $H_0$ of about 10\% with current generation GW detectors with a configuration similar to our HLVK. Our sensitivity results are, instead, somewhat in contradiction with \cite{Mukherjee:2022afz}, which performs a cross-correlation analysis using the BBH events available at the time, about 10, with galaxy catalogs finding a $\sim 20\%$ determination of the Hubble constant, while for the same results we find that several thousand events are required. The reason of the discrepancy is unclear.

\subsection{ET2CE Results}

Thanks to its very large statistics of GW events with precise localization, the ET2CE configuration is able to provide much stronger constraints with respect to  HLVK and HLVIK. We can see from  \autoref{fig:components} and \autoref{tab:h_results} that  XC  gives a sensitivity to $h$ of about 1\% as opposed to 4\% for GG. 
Put in another way, the information coming from the \emph{pure} cross-correlation between the galaxy catalogue and a GW catalogue is able to provide a 1\% sensitivity on $h$. 

Given the current tension on the determination of this parameter coming from different measurements, an independent determination at the 1\% level will be crucial to shed light on the issue.
Furthermore, when GG and XC information are considered together, the sensitivity improves to 0.7\%.

We show in \autoref{fig:ET_2CE_Triangle} the one-dimensional posteriors and 
the two-dimensional contours for the three ET2CE forecasts, namely GG, XC, and full information. Also, in this case, we do not report the WW case since it provides only extremely loose constraints.
We see that only in the case of the Hubble parameter,  XC is able to provide better constraints than GG.
This is not surprising since GWs behave as standard sirens, and thus are particularly sensitive to $h$.
For other parameters,  GG  is more constraining than XC, but  XC  is still able to provide significant constraints. Most importantly, the parameter degeneracy directions of  GG and XC  tend to be orthogonal to each other. Thus, when the GG,  XC and WW information are considered altogether, the constraints are significantly stronger than with any of the probes considered alone. This is particularly clear when one looks at the $h-\omega_{\rm b}$ and $h-\omega_{\rm cdm}$ contour planes. The ability of this technique to break degeneracies demonstrates its importance for future cosmological analyses.

Finally, \autoref{fig:ET_2CE_Triangle} shows that the XC probe alone is not able to provide significant constraints on the GW bias normalisation parameter ($a_1^W$) due to various degeneracies with other cosmological parameters, while the constraint becomes very tight when the information from GG is used jointly. This result is in agreement with similar analyses specifically focused on the study of the GW bias, like in \cite{Calore:2020bpd}, with the caveat that in \cite{Calore:2020bpd} the cosmology was kept fixed. The present analysis thus complements these studies and highlights the fact that in order to constrain the GW bias effectively,
one needs to use additional information beyond the GW$\times$galaxy cross-correlation alone.

The numerical results for the various cosmological parameters for the case shown in \autoref{fig:ET_2CE_Triangle}, are presented in \autoref{app:numresults}. The tight constraints obtained from this analysis confirm the agnostic nature of our priors in \autoref{tab:cosmo_nuisance}.

\begin{table}[t]
  \centering

  \begin{tabular}{|c||c|c|}
  \hline
    \multicolumn{3}{|c|}{\textbf{ET2CE}} \\
    \hline
    \hline
    \textbf{Population} & \textbf{Contribution} & \makecell{\textbf{$H_0$ bestfit$_{-1\sigma}^{+1\sigma}$} \\ 
    $\left[{\rm km}\,{\rm s}^{-1}\,{\rm Mpc}^{-1}\right]$} \\
    \hline
    \hline
    \makecell{\textbf{Galaxy}\\\textbf{Catalogue}}
    & GG & $67.4_{-2.8}^{+2.3}$ \\
    \hline
    \hline
    \multirow{2}{*}{\textbf{BBH}} 
    & XC & $67.4_{-1.2}^{+1.2}$ \\
    & Full matrix &  $67.3_{-0.5}^{+0.5}$\\
    \hline
    \hline
    \multirow{2}{*}{\textbf{BNS}}    
    & XC & $67.4_{-1.3}^{+1.5}$ \\
    & Full matrix &  $67.4_{-0.8}^{+0.8}$\\
    \hline
    \hline
    \multirow{2}{*}{\textbf{BBH+BNS}}    
    & XC & $67.3_{-0.9}^{+0.8}$\\
    & Full matrix & $67.2_{-0.4}^{+0.5}$ \\
    \hline
  \end{tabular}
  \caption{Hubble constant $H_0$ best-fit and error for the ET2CE case considering BNS and BBH dark sirens samples individually.  The fiducial value of $H_0$ is 67.4.}
  \label{tab:h_results_ET_2CE}
\end{table}

\begin{figure}[ht]
    \hspace{0.1\textwidth}
    \includegraphics[height=0.35\textwidth]{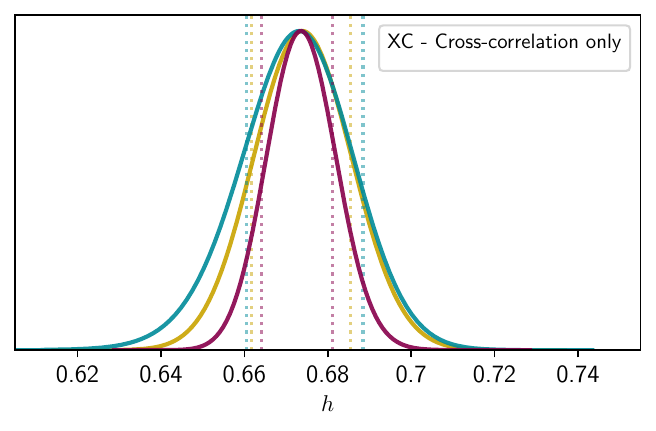}
    \includegraphics[height=0.35\textwidth]{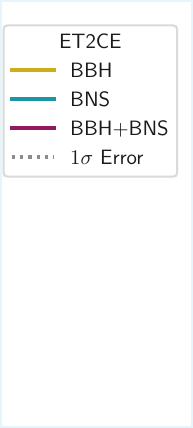}
    
    \hspace{0.1\textwidth}
    \includegraphics[height=0.35\textwidth]{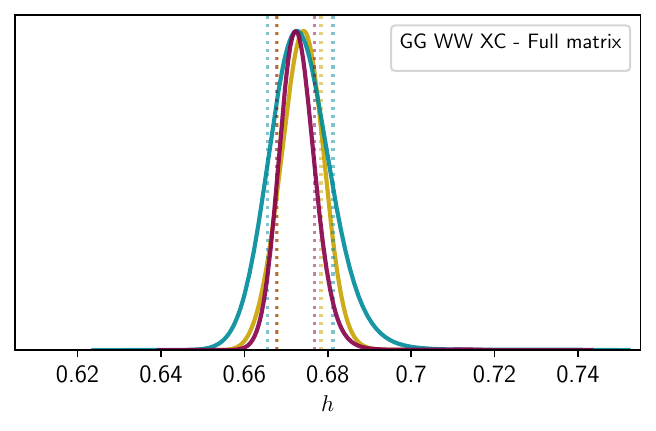}
    \caption{Posterior on the Hubble constant $h$ for the ET2CE configuration. The top plot shows the XC case, and the bottom one the full matrix case. In each plot, the separate BBH and BNS cases are shown together with the joint BBN+BBH case.  The dotted vertical lines show the $1\sigma$ intervals, which are also reported numerically in \autoref{tab:h_results_ET_2CE}. }
    \label{fig:ET_2CE_binaries}
\end{figure}

\subsubsection{ET2CE BBH and BNS break-down}

The results of the previous section refer to the case in which BBH and BNS data are considered together. 
Results for the BBH-only, BNS-only and combined cases are shown in \autoref{fig:ET_2CE_binaries} and \autoref{tab:h_results_ET_2CE}.
As expected from the not-too-different SNRs calculated in \autoref{tab:SNR}, BBH and BNS provide similar constraints on $h$ when considering only the XC probe. The BNS constraints are only slightly worse than the BBH ones. 
This is confirmed in the full information case (GG WW XC), where again the BNS-only data gives similar but slightly worse constraints than the BBH-only data. The combined case is very similar to the BBH-only case.

\section{Summary and Conclusions}\label{sec:conclusion}

Modelling galaxies and dark sirens as two linear biased tracers of the underlying dark matter density field, we have evaluated the amount of cosmological information contained in the cross-correlation between future observable maps of galaxies and gravitational waves (GWs). We were able to confirm that this technique is promising, in particular, as a way to measure the Hubble expansion rate independently of standard candles (like type Ia supernovae) or standard rulers (like the sound horizon).

Contrary to previous works on the subject, we perform a full likelihood forecast, i.e., we fit mock data to a cosmological model (specifically the flat $\Lambda$CDM model) and the bias of each tracer, and we sample the parameter space with Monte Carlo Markov Chains (MCMC). We use a tomographic approach: we divide the galaxy catalogue into redshift bins and the dark siren catalogue into luminosity distance bins. The fact that one can only measure redshifts for galaxies and luminosity distances for dark sirens is not an issue in the context of a Bayesian analysis. In such an analysis, at each point in parameter space, one needs to assume a given cosmological model. Then, the luminosity-distance-to-redshift relation is known,  and it is straightforward to express everything in redshift space. Thus, for this cosmology, one can compute the cross-correlation spectrum between the galaxy and GW maps and evaluate the likelihood of the data given the model. The most likely cosmology is the one in which the anisotropies in the two catalogues overlap at each redshift, such that the observed cross-correlation matches theoretical predictions.

Regarding galaxies, our study assumes the sensitivity of forthcoming photometric galaxy surveys like Euclid and LSST. For dark sirens, we considered the binary black holes (BBH) and binary neutron stars (BNS) expected to be detectable with a combination of either existing detectors (LIGO, Virgo, KAGRA) or planned third-generation detectors (Einstein Telescope, Cosmic Explorers). More specifically, we considered three configurations, which we dubbed HLVK, HLVIK and ET2CE, in all three cases for a data-taking period of 10 years. We reach the following conclusions:
\begin{itemize}
  \item The HLVK configuration has limited capabilities to accurately map GW anisotropies and measure their cross-correlation with a (photometric) galaxy survey. It can only constrain the Hubble constant to about 30\%.
  The HLVIK configuration, where LIGO India is added to the network, performs significantly better, providing constraints on $H_0$ of the order of 7\%, which is similar to what is achievable using the auto-correlation of galaxy maps from the same survey.

 \item Instead, the cross-correlation between the network of 3G detectors ET2CE and galaxy data is able to provide tight constraints on the Hubble constant, even after marginalising over unknown dark siren bias parameters. This cross-correlation has a sensitivity to $H_0$ of 1\%, about four times better than the auto-correlation of galaxy maps. 

  \item The cross-correlation data alone cannot resolve degeneracies between the dark siren bias parameters and the cosmological parameters (except $H_0$). A combination with the galaxy auto-correlation data resolves these degeneracies and allows us to tightly constrain the GW biases and extract additional information on all cosmological parameters from the cross-correlation data.
  
 \item We further find that the directions of degeneracy between $H_0$ and other cosmological parameters are orthogonal for the cross-correlation and auto-correlation. Thus, when the two probes are combined,  constraints are stronger than from each probe alone, in particular for $\omega_b$ and $\omega_{cdm}$. $H_0$ constraints, instead, are dominated by the cross-correlation. Thus, adding the auto-correlation information provides only a mild improvement (from 1\%  to 0.7\%).

 \item Finally, we find that BBHs and BNSs provide similar constraints, although BBHs score slightly better, as expected due to their better event reconstruction (positional and distance) properties.

\end{itemize}

There exist other methods to extract cosmological information from dark sirens, like the galaxy catalogue association technique \cite{Gray:2023wgj,LIGOScientific:2019zcs,LIGOScientific:2021aug, Beirnaert2025,Gair:2022zsa, Borghi_2024} or the spectral sirens approach \cite{spectralsirens,2025arXiv250904348T,MaganaHernandez:2025cnu} method. These methods sometimes predict a better performance than the cross-correlation technique investigated here. However, we stress that the latter is very robust and provides basically model-independent results, not relying, for instance, on the completeness of the catalogue used, nor on the assumed dark siren mass distribution. We thus highlight the importance of this technique for future cosmological analyses using GWs.


\acknowledgments
We are very grateful to Joline Noltmann for developing a preliminary version of a dark siren likelihood during her Bachelor thesis.
We wish to thank Raul Abramo and Pasquale Serpico for reading the manuscript and providing useful comments. 
AC acknowledges support from the Research grant TAsP (Theoretical Astroparticle Physics) funded by INFN, and Research grant ``Addressing systematic uncertainties in searches for dark matter'', Grant No.\ 2022F2843L, CUP D53D23002580006 funded by the Italian Ministry of University and Research (\textsc{mur}). GS and JL acknowledge support from the DFG. 
Calculations were conducted on the Lichtenberg high-performance computer of the TU Darmstadt, with computing resources granted by RWTH Aachen University under
project ‘rwth1661’ and ‘rwth1695’.
\bibliographystyle{JHEP.bst}
\bibliography{TheBib}

\appendix

\section{Extension to spatially curved universe}\label{app:dl_z}

Our likelihood would be straightforward to extend to other cosmologies, including models with spatial curvature. In that case, the luminosity distance would read
\begin{equation}
    \centering\label{eq:D_L-def}
    D_L = a(t_0)(1+z)f_\kappa\left(\int_{0}^{z}\frac{cdz'}{a(t_0)H(z')} \right)~,
\end{equation}
with
\begin{equation}\centering\label{eq:comoving_distance_curvature}
    f_\kappa(r) =
\begin{cases}
  \frac{1}{\sqrt{\kappa}}\sin{\sqrt{\kappa}r}, & \text{if } \kappa > 0 \\
  r, & \text{if } \kappa = 0 \\
  \frac{1}{\sqrt{-\kappa}}\sinh{\sqrt{-\kappa}r}, & \text{if } \kappa < 0~. \\
\end{cases}
\end{equation}
Then, \eqref{eq:dDLdz_flat} would need to be generalised as
\begin{equation}
    \centering\label{eq:dldz}
    \frac{dD_L}{dz} = \frac{D_L}{1+z}+\frac{1+z}{H(z)}\left.\frac{df_\kappa(r)}{dr}\right|_{\bar{r}}~,
\end{equation}    
with
\begin{equation}
    \bar{r} =\int_{0}^{z}\frac{cdz'}{a(t_0)H(z')}~.
\end{equation}

\section{Examples of $C_\ell$}\label{app:cls}

\begin{figure}[ht]
    \centering
    \begin{subfigure}[b]{0.5\textwidth}
        \centering
        \includegraphics[width=\textwidth]{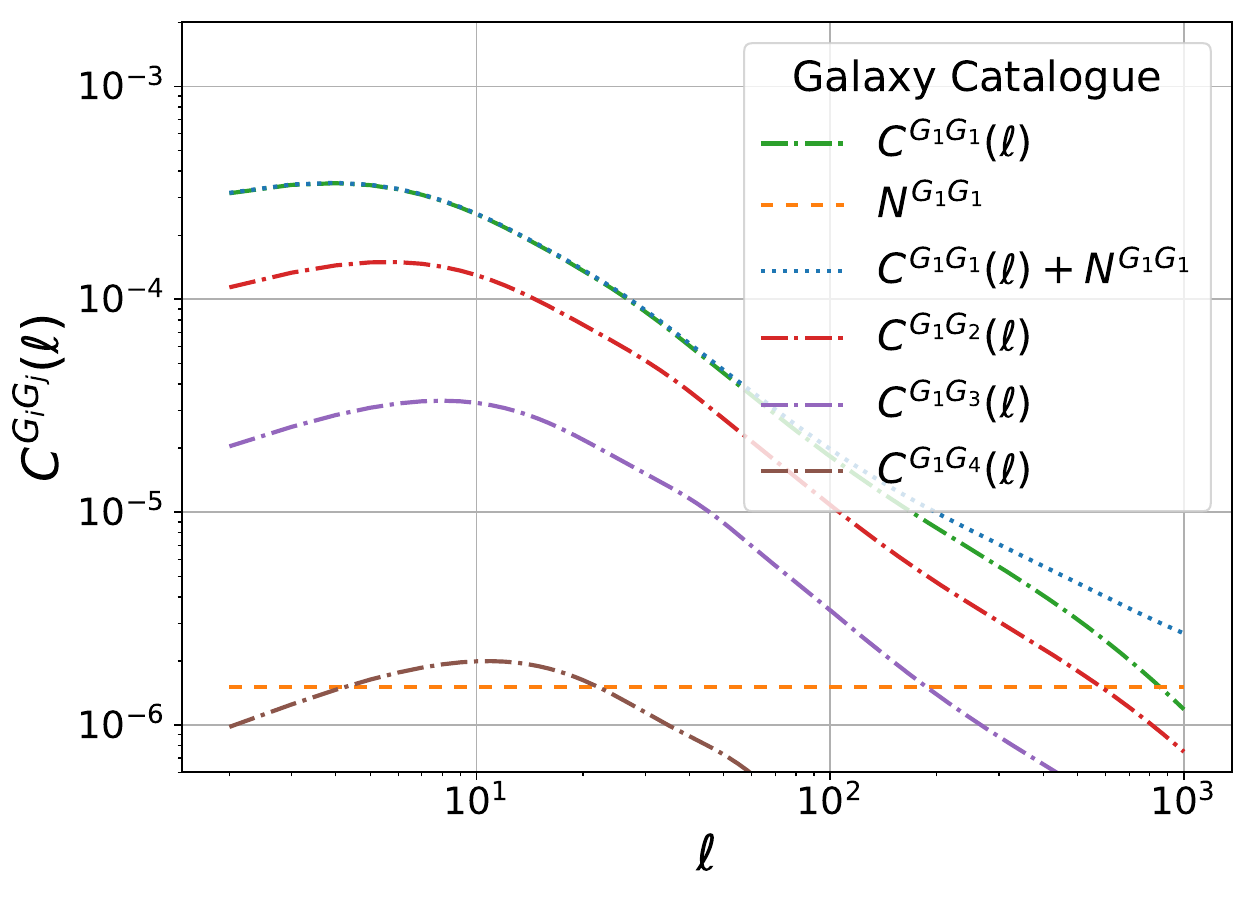}
        \caption{Galaxy auto-correlation $C_\ell$}
        \label{fig:ClGG}
    \end{subfigure}
    \hfill
    \begin{subfigure}[b]{0.5\textwidth}
        \centering
        \includegraphics[width=\textwidth]{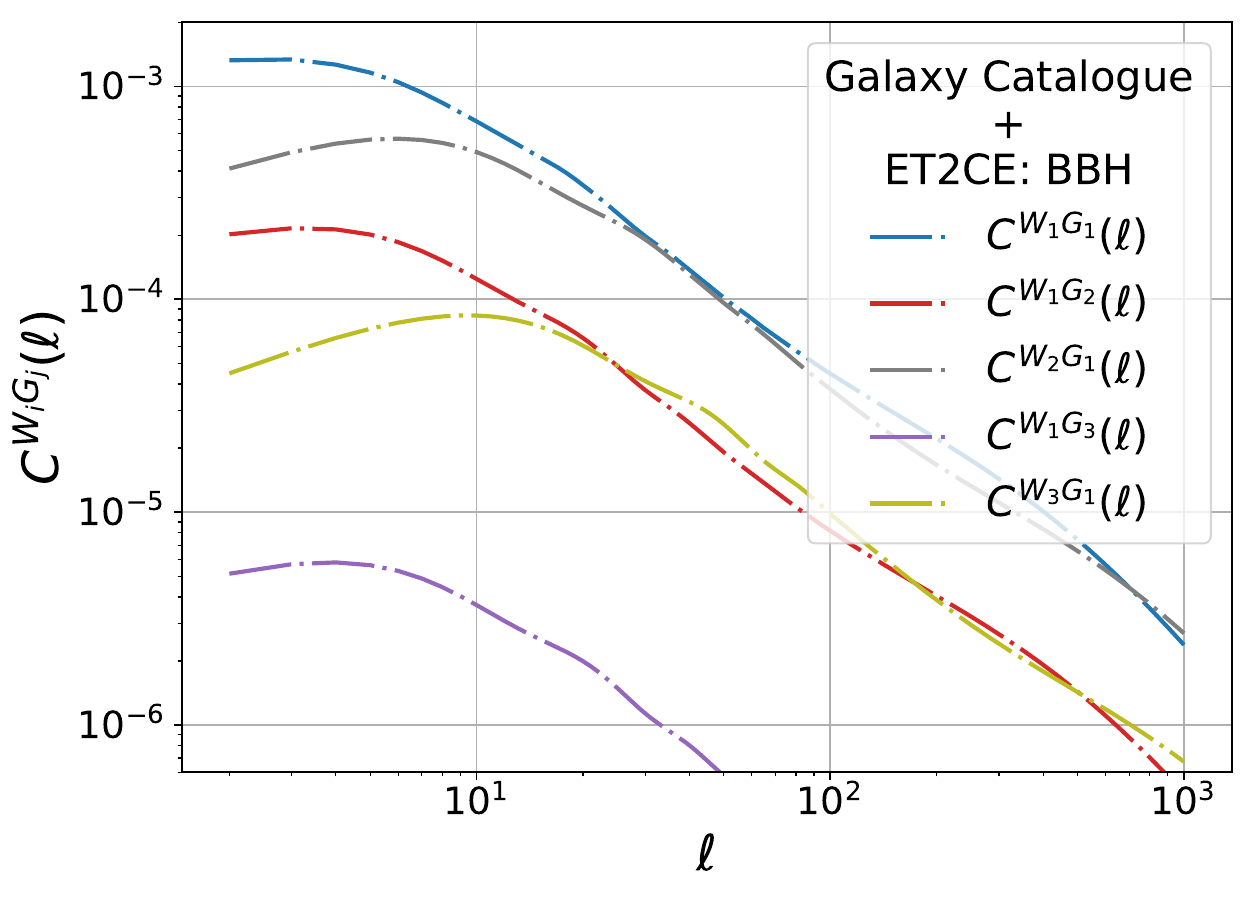}
        \caption{Dark sirens-galaxies cross-correlation $C_\ell$}
        \label{fig:ClWG}
    \end{subfigure}
    \hfill
    \begin{subfigure}[b]{0.5\textwidth}
        \centering
        \includegraphics[width=\textwidth]{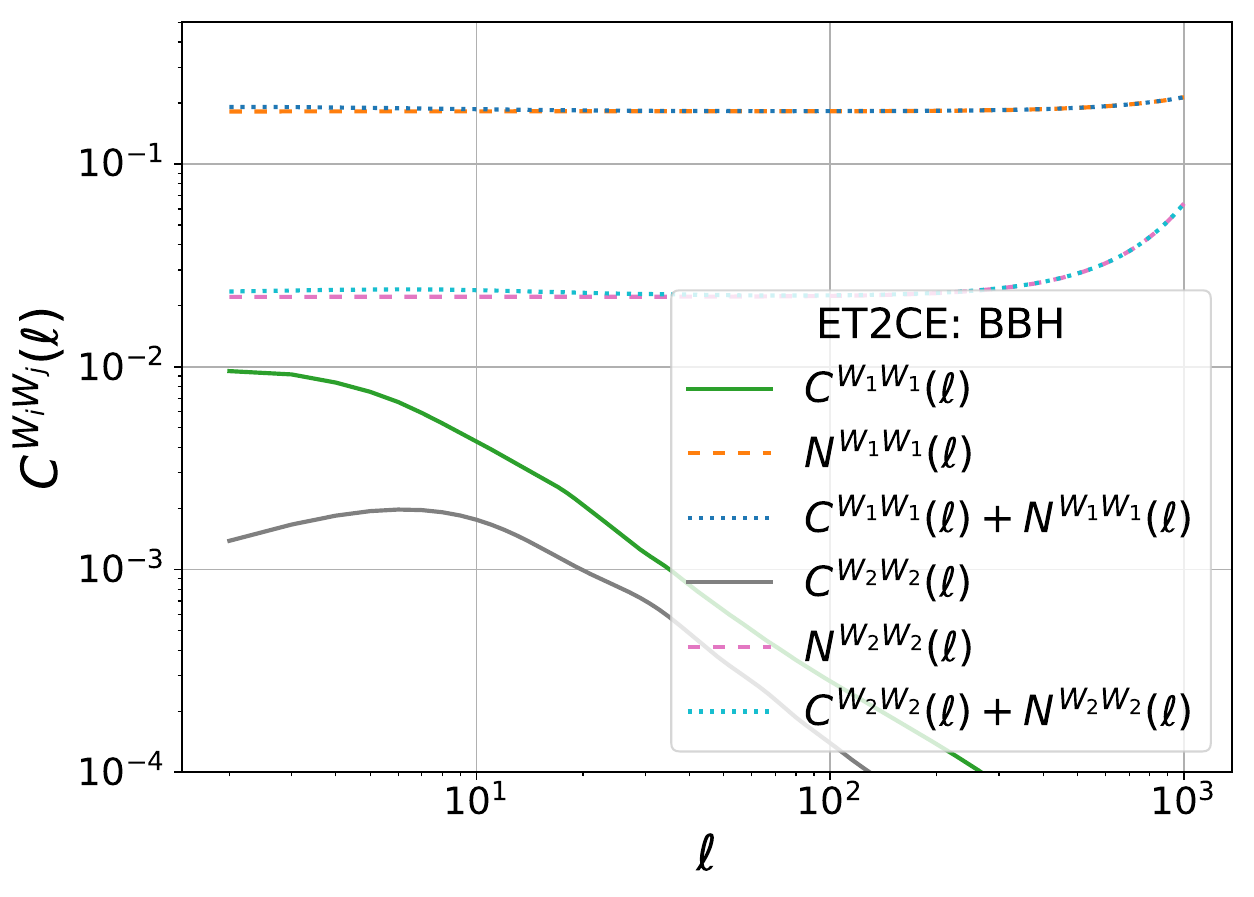}
        \caption{Dark sirens auto-correlation $C_\ell$}
        \label{fig:ClWW}
    \end{subfigure}
    \caption{Representative angular power spectrum components for the observation of BBH with the ET2CE configuration. ${C}_\ell$ and $N$ represent respectively the noiseless angular power spectrum \eqref{eq:C_general} and the noise term \eqref{eq:noise_def}.}
    \label{fig:Cl}
\end{figure}

Following the discussion of \autoref{sec:power}, \autoref{fig:Cl} shows some representative cases of the $C_\ell$s described by equations (\ref{eq:Cl_GG}-\ref{eq:Cl_WW}). More specifically, the case of BBH observations with the ET2CE configuration is considered. Galaxy clustering and dark sirens auto-correlation, reported in \autoref{fig:ClGG} and \ref{fig:ClWW}, show the interplay between signal and noise, which is only present in the auto-correlation of each redshift bin. Note the large noise dominating the GW signal in the bottom plot, highlighting the importance of the cross-correlation for extracting new information from GW.

\section{Numerical constraints}\label{app:numresults}
The full set of numerical results for the constraints coming from the ET2CE analysis with only the BBH catalogue, illustrated  in \autoref{fig:ET_2CE_Triangle}, is reported in \autoref{tab:results_ET_2CE}.
The result also shows that the priors are not informative.

{
\renewcommand{\arraystretch}{1.5}
\begin{table}[h]
  \centering

  \begin{tabular}{|c||c||c|c c|}
  \cline{2-5}
    \multicolumn{1}{c}{} & \multicolumn{4}{|c|}{\textbf{ET2CE}} \\
    \cline{2-5}
    \noalign{\vskip 2pt}
    \cline{2-5}
   \multicolumn{1}{c}{} &\multicolumn{1}{|c||}{\textbf{Population}} & \makecell{\textbf{Galaxy}\\\textbf{Catalogue}} & \multicolumn{2}{|c|}{\textbf{BBH}} \\
    \cline{2-5}
    \noalign{\vskip 2pt}
    \cline{2-5}
    \multicolumn{1}{c}{} & \multicolumn{1}{|c||}{\textbf{Contribution}} 
    & GG & XC & Full matrix\\
    \cline{2-5}
    \noalign{\vskip 4pt}
    \hline 
    \multirow{8}{*}{
    \makecell{\textbf{Parameters: }\\\textbf{bestfit$_{-1\sigma}^{+1\sigma}$}}}
     &  $\omega_{\rm b}$ & $2.24_{-0.32}^{+0.24}\times 10^2$ 
     &  $2.37_{-0.68}^{+0.79}\times 10^2$ &  $2.23_{-0.08}^{+0.08}\times 10^2$ \\
      & $\omega_{\rm cdm}$ & $0.120_{-0.007}^{+0.005}$ 
      & $0.114_{-0.010}^{+0.010}$ & $0.116_{-0.001}^{+0.001}$ \\
      & $n_{\rm s}$ &  $0.966_{-0.007}^{+0.007}$ 
      &  $0.98_{-0.11}^{+0.08}$ &  $0.966_{-0.004}^{+0.004}$ \\ 
      & $A_{\rm s}$ &  $2.42_{-0.09}^{+0.09}\times 10^{-9}$ 
      & $2.45_{-0.64}^{+0.43}\times 10^{-9}$ &  $2.43_{-0.04}^{+0.04}\times 10^{-9}$\\
        & $h$   &  $0.67_{-0.03}^{+0.02}$   
      & $0.674_{-0.012}^{+0.012}$   &  $0.673_{-0.005}^{+0.005}$ \\
      \cline{2-5}
      & $a_1^G$ &  $1.00_{-0.01}^{+0.01}$ 
      & $0.87_{-0.62}^{+0.18}$ &  $1.00_{-0.01}^{+0.01}$ \\
      & $a_1^W$  &   
      & $2.34_{-1.13}^{+0.52}$ &  $2.00_{-0.13}^{+0.11}$ \\
      & $a_2^W$  &   
      & $-0.01_{-0.28}^{+0.73}$ &  $-0.10_{-0.59}^{+0.57}$\\
    \hline
  \end{tabular}
  \caption{Best-fit and error for the ET2CE case considering BBH dark sirens samples for all cosmological parameters, also illustrated in \autoref{fig:ET_2CE_Triangle}. The fiducial values and priors are reported in \autoref{tab:cosmo_nuisance}.}
  \label{tab:results_ET_2CE}
\end{table}
}

\section{Triangle plots}\label{app:triangle}

\begin{figure}[hp]
    \includegraphics[width=\textwidth]{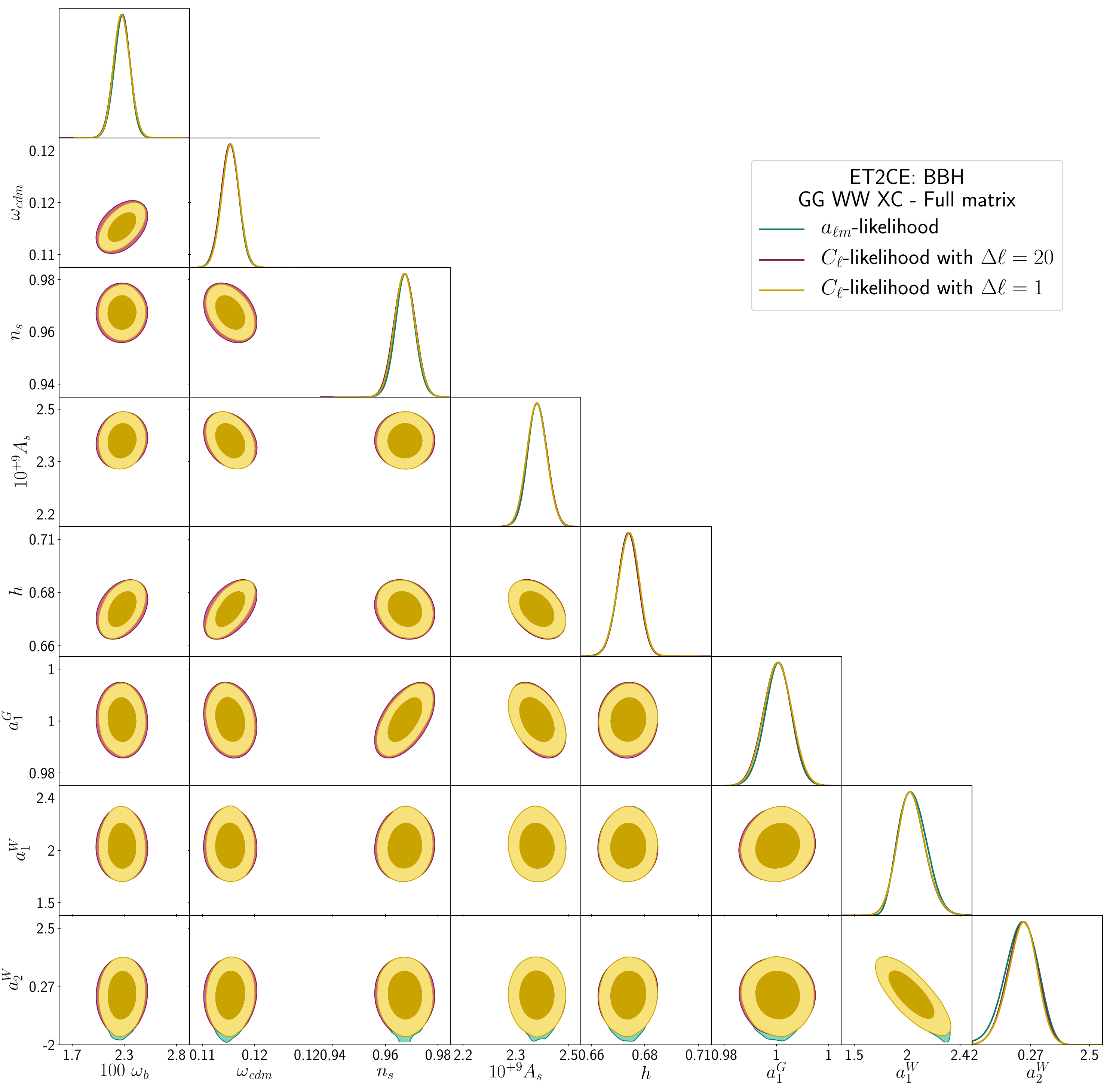}
    \caption{Comparison of the results arising from different likelihoods for the ET2CE configuration with BBH as dark sirens source. The three cases shown are the $a_{\ell m}$-based likelihood, a $C_\ell$-based likelihood linearly binned in $\ell$-space with $\Delta \ell=20$, and $C_\ell$-based likelihood with $\Delta \ell=1$ used in this work. The full triangle plot is shown. We can see that the three cases are in very good agreement.}
    \label{fig:alm_bin_Triangle}
\end{figure}

Additional triangle plots of the cosmological parameters that extend the discussion in \autoref{sec:likelihood} and \ref{sec:results} are reported in this appendix.  \autoref{fig:alm_bin_Triangle} presents a comparison of the results from MCMC runs with three different likelihoods.
The first set of results comes from an $a_{\ell m}$-based likelihood, following the formalism from  \cite{Audren:2012vy}. The same analysis is then repeated with the $C_\ell$-based likelihood in two cases, binning the $\ell$-space with $\Delta\ell=20$ and without binning (or $\Delta\ell=1$). The latter configuration is the one used for the results of this work.
\autoref{fig:alm_bin_Triangle} shows good agreement between the three methods.

\begin{figure}[hp]
    \includegraphics[width=\textwidth]{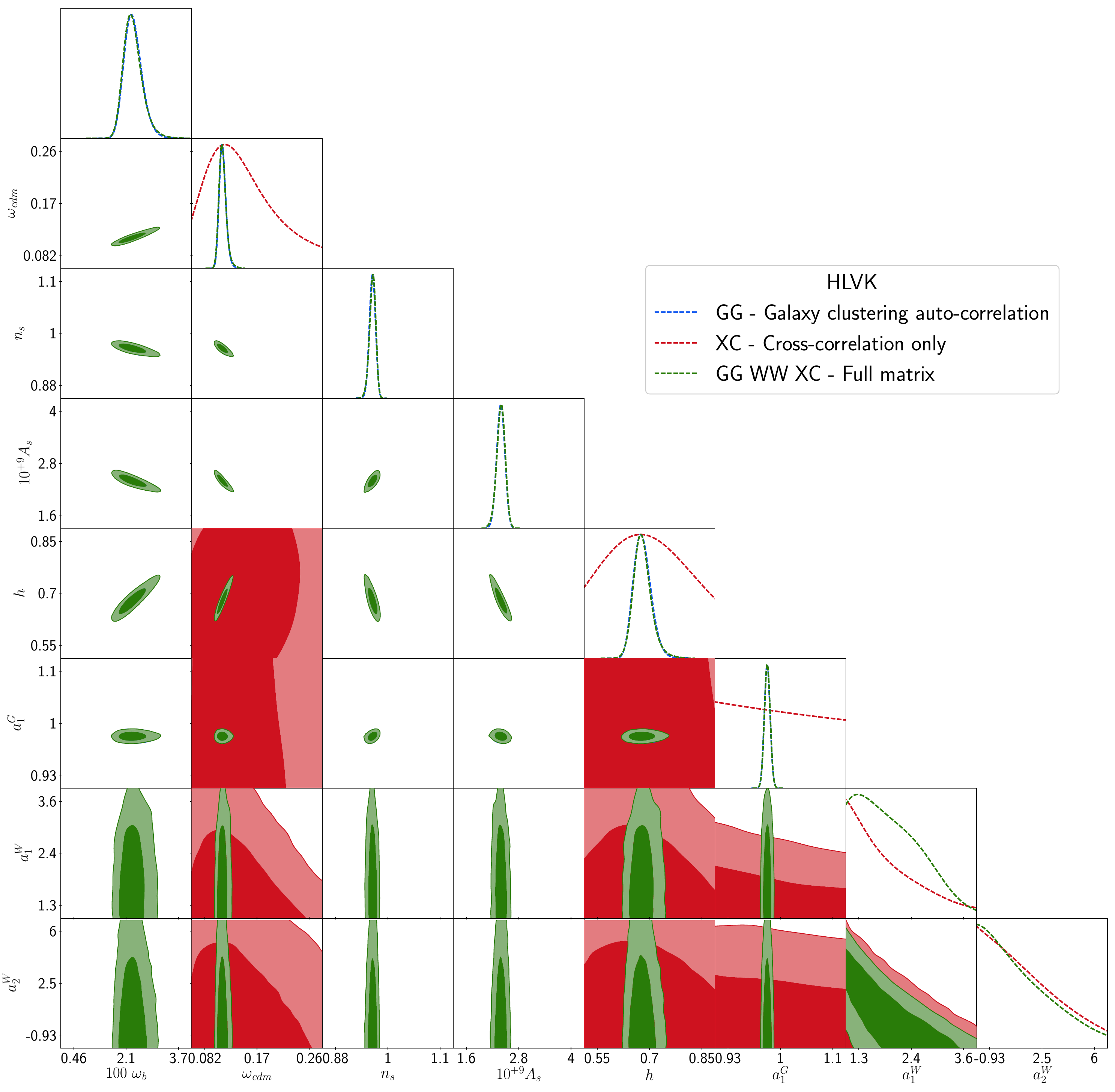}
    \caption{Full triangle plot for MCMC scans for the GG, XC, and full matrix cases considering BBH detections with HLVK. $1\sigma$ and $2\sigma$ contours are shown. Note that  $\omega_b$, $n_s$ and $A_s$ have been kept fixed for the XC cross-correlation scan, and are thus not shown in that case. }
    \label{fig:LVK_Triangle}
\end{figure}

\begin{figure}[hp]
    \includegraphics[width=\textwidth]{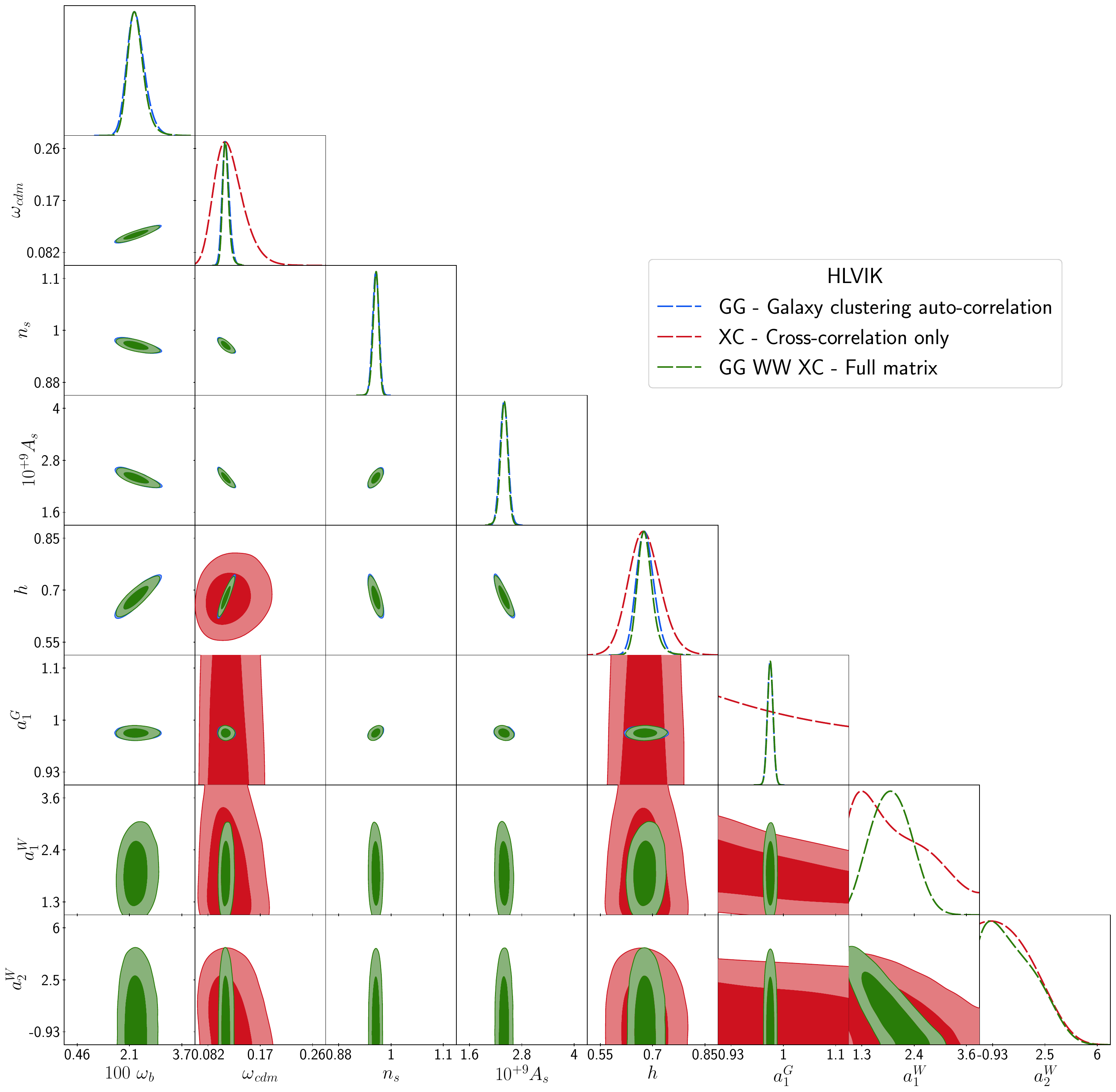}
    \caption{Full triangle plot for MCMC scans for the GG, XC, and full matrix cases considering BBH detections with HLVIK. $1\sigma$ and $2\sigma$ contours are shown. 
    Note that  $\omega_b$, $n_s$ and $A_s$ have been kept fixed for the XC cross-correlation scan, and are thus not shown in that case. }
    \label{fig:HLVIK_Triangle}
\end{figure}

\autoref{fig:LVK_Triangle} and \ref{fig:HLVIK_Triangle} are analogous to \autoref{fig:ET_2CE_Triangle} in the main text, and report the full triangle plot for the HLVK and HLVIK cases, respectively. Due to the small dark sirens sample and reduced constraining power compared to the ET2CE case, the parameters $\omega_b$, $n_s$ and $A_s$ have been kept fixed to the benchmark values in the pure cross-correlation XC  MCMC scan.

Looking  at \autoref{fig:ET_2CE_Triangle} together with \autoref{fig:LVK_Triangle} and \ref{fig:HLVIK_Triangle} we can see that for the ET2CE full-matrix case the posterior of the parameters is to a good approximation Gaussian. In this case our full-likelihood approach and Fisher matrix-based approaches are expected to return comparable results.
Deviation from Gaussianity, especially in the form of elongated degeneracies, instead,  appear for the ET2CE-XC case and both HLVK and HLVIK cases, in particular in the bias-related parameters. In these cases, the full-likelihood  approach is particularly important and Fisher matrix computations are expected to be only approximate and miss some of the degeneracy directions.

\end{document}